\documentclass[apj,twocolumn]{emulateapj}
\newcommand{\lsim}{\raisebox{-0.13cm}{~\shortstack{$<$ \\[-0.07cm] $\sim$}}~}
\newcommand{\gsim}{\raisebox{-0.13cm}{~\shortstack{$>$ \\[-0.07cm] $\sim$}}~}
\newcommand{\ffm}{4.5 \, \rm  \mu m}
\newcommand{\tfm}{24 \, \rm  \mu m}
\newcommand{\ksau}{K_s^{\rm auto}>24}

\shorttitle{{\em Spitzer} bright, UltraVISTA faint sources in COSMOS}
\shortauthors{Caputi et al.}

\begin{document}

\title{{\em Spitzer} bright, UltraVISTA faint sources in COSMOS:  the contribution to the overall population of massive galaxies at $\lowercase{z}=3-7$}


\author{K. I. Caputi\altaffilmark{1},
O. Ilbert\altaffilmark{2},
C. Laigle\altaffilmark{3},
H. J. McCracken\altaffilmark{3},
O. Le F\`evre\altaffilmark{2},
J. Fynbo\altaffilmark{4},
B. Milvang-Jensen\altaffilmark{4},
P. Capak\altaffilmark{5},
M. Salvato\altaffilmark{6},
Y. Taniguchi\altaffilmark{7}
}


\altaffiltext{1}{Kapteyn Astronomical Institute, University of Groningen, P.O. Box 800, 9700 AV Groningen, The Netherlands.\\ Email: karina@astro.rug.nl}
\altaffiltext{2}{Aix Marseille Universit\'e, CNRS, LAM (Laboratoire d'Astrophysique de Marseille), UMR 7326, 13388, Marseille, France}
\altaffiltext{3}{Sorbonne Universit\'es, UPMC Universit\'e Paris 6 et CNRS, UMR 7095, Institut d'Astrophysique de Paris, 98 bis Boulevard Arago, 75014 Paris, France}
\altaffiltext{4}{Dark Cosmology Center, Niels Bohr Institute, University of Copenhagen, Juliane Maries Vej 30, 2100 Copenhagen, Denmark}
\altaffiltext{5}{Spitzer Science Center, California Institute of Technology, Pasadena, CA, 91125, USA}
\altaffiltext{6}{Max-Planck-Institut f\"ur Extraterrestrische Physik, Postfach 1312, 85741, Garching bei M\"unchen, Germany; Excellence Cluster, Boltzmann Strasse 2, 85748, Garching, Germany}
\altaffiltext{7}{Research Center for Space and Cosmic Evolution,
Ehime University, 2-5 Bunkyo-cho, Matsuyama 790-8577, Japan}

\begin{abstract}
We have analysed a sample of 574 {\em Spitzer} $4.5 \, \rm \mu m$-selected galaxies with $[4.5]<23$  and $\ksau$~(AB) over the UltraVISTA ultra-deep COSMOS field. Our aim is to investigate whether these mid-IR bright, near-IR faint sources contribute significantly to the overall population of massive galaxies at redshifts $z \geq 3$. By performing a spectral energy distribution (SED) analysis using up to 30 photometric bands, we have determined that the redshift distribution of our sample peaks at redshifts $z \approx 2.5-3.0$, and $\sim32\%$ of the galaxies lie at $z \geq 3$.  We have studied the contribution of these sources to the galaxy stellar mass function (GSMF) at high redshifts. We found that the $[4.5]<23$, $\ksau$ galaxies produce a negligible change to the GSMF previously determined for $K_s^{\rm auto}< 24$ sources at $3 \leq z <4$, but their contribution is more important at $4 \leq z <5$, accounting for $\gsim 50\%$ of the galaxies with stellar masses $M_{\rm st} \gsim 6 \times 10^{10} \, \rm M_\odot$. We also constrained the GSMF at the highest-mass end ($M_{\rm st} \gsim 2 \times 10^{11} \, \rm M_\odot$) at $z \geq 5$.  From their presence at $5 \leq z<6$, and virtual absence at higher redshifts, we can pinpoint quite precisely the moment of appearance of the first most massive galaxies as taking place in the $\sim 0.2 \, \rm Gyr$ of elapsed time between $z\sim6$ and $z\sim5$. Alternatively, if very massive galaxies existed earlier in cosmic time, they should have been significantly dust-obscured to lie beyond the detection limits of current, large-area, deep near-IR surveys.

\end{abstract}


\section{Introduction}
\label{sec-intro}

Constraining the number density of massive ($M_{\rm st} \gsim 5 \times 10^{10} \, \rm M_\odot$) galaxies at different redshifts is very important to understand when galaxy buildup proceeded most efficiently in cosmic time. Over the past decade, multiple studies have shown that a significant fraction of the massive galaxies that we know today were already in place and massive at $z\sim2$  (e.g. Saracco et al.~2004, 2005; Caputi et al.~2005, 2006; Labb\'e et al.~2005; Pozzetti et al.~2007). The search for massive galaxies has also been extended to higher redshifts $z\sim 3-4$, usually through the analysis of sources with red near-infrared (near-IR) colours that were mostly undetected at optical wavelengths \citep[e.g.][]{fra03,kod07,rod07}. Until recently, however, the typical depths of deep near-IR surveys ($K_s\sim 23.5-24.0$ AB mag) and small areas covered  have prevented a systematic search for the rare massive galaxies present at $ z \gsim 4$.

The first epoch of appearance of massive galaxies in the early Universe  constitutes an important constraint for galaxy formation models. These models predict that massive galaxies are formed in the high-density fluctuation peaks of the matter density field \citep{col89,mo96}. Reproducing the number density of massive galaxies since redshift $z\sim7$ is critical to explain how massive galaxy formation proceeded after the epoch of reionisation, and the subsequent galaxy buildup until today.

The galaxy stellar mass function (GSMF) is a very important tool to investigate the number density evolution of galaxies of different stellar masses through cosmic times. Optical galaxy surveys have enabled the study of the GSMF up to redshift $z\sim1$ \citep[e.g.][]{poz10,bal12,dav13,mou13}, while near-IR surveys have extended these studies up to $z\sim3-4$  \citep{fon06,kaj09,mar09,bie12,ilb13,muz13}. Recent ultra-deep near-IR surveys over small areas of the sky have allowed for GSMF studies at higher redshifts, providing constraints at the intermediate stellar-mass regime, i.e., $M_{\rm st} \sim  10^9 - 10^{10} \, \rm M_\odot$ \citep[e.g.][]{gon11,san12,dun14,gra15}. However, the small surveyed areas make it difficult to properly sample the GSMF high-mass end, as massive galaxies are rare at high $z$. Alternatively, galaxy selections based on mid-IR images taken with the {\em Spitzer Space Telescope} Infrared Array Camera \citep[IRAC;][]{faz04},  some of which were conducted over larger areas of the sky, have offered the possibility of exploring the GSMF high-mass end up to $z\sim5$ \citep{pg08,man09,ilb10,cap11,dsz11,ste15}.

The depth of the near-IR maps typically available in these larger fields, namely $K_s \sim 23.5-24.0$,  was sufficient to identify the vast majority ($\sim95-98\%$) of the  IRAC sources to $[4.5]=23$~AB~mag. The small percentage of sources that remained unidentified were usually neglected, as it was impossible to derive any of their properties --including their redshifts-- without any detection beyond the IRAC bands. Indeed, if these unidentified sources had a similar redshift distribution to those that are identified, then taking them into account would not make any significant difference to the already derived results, in particular those on the number density of massive galaxies at high $z$. However, if these missing galaxies had a {\em biased redshift distribution} towards high redshifts, their contribution in the early Universe could be more important than previously assumed.

Hints on the importance of mid-IR bright, near-IR faint sources towards the study of massive galaxies at high redshifts have been provided by the analysis of IRAC extremely red sources \citep{wik08,hua11,cap12}, which showed that most of them are massive galaxies at $3 \lsim z \lsim 5$. However, these examples are extreme cases of the mid-IR bright, near-IR faint galaxies that are present in ultra-deep near-IR maps. These sources are not fully representative of the entire population of bright IRAC sources that remain beyond the typical identification limits of wide-area, near-IR surveys (i.e., $K_s\sim24.0$). 

In this paper we investigate a more representative sample of the IRAC bright sources that have been unidentified so far in large-area, deep near-IR surveys. Our aim is understanding their importance within the overall population  of massive galaxies at high redshifts ($z \geq 3$). This study is possible thanks to the on-going UltraVISTA near-IR survey \citep{mcc12}, which at the current stage already has a unique combination of area and photometric depth. The layout of this paper is as follows. In Section \S\ref{sec-data} we describe the photometric datasets used. In Section \S\ref{sec-sample}, we explain our sample selection, and photometric redshift determinations through a spectral energy distribution (SED) analysis. In Sections \S\ref{sec-massz35} and \S\ref{sec-massz57}, we present our results for galaxies at $3 \leq z < 5$ and $5 \leq z < 7$, respectively, including their contributions to the GSMF at these redshifts. We conduct an updated analysis of the cosmic stellar mass density evolution in Section \S\ref{sec-cstmd}. Finally, in Section \S\ref{sec-conc}, we summarize our findings and give some concluding remarks. We adopt throughout a cosmology with $\rm H_0=70 \,{\rm km \, s^{-1} Mpc^{-1}}$, $\rm \Omega_M=0.3$ and $\rm \Omega_\Lambda=0.7$. All magnitudes and fluxes are total, and refer to the AB system \citep{oke83}, unless otherwise stated. Stellar masses correspond to a Salpeter IMF over (0.1-100)~$\rm M_\odot$.

\section{Datasets}
\label{sec-data}

The Cosmic Evolution Survey  \citep[COSMOS; ][]{sco07} comprises a wealth of multi-wavelength imaging and spectroscopic data covering $\sim 1.4 \times 1.4 \, \rm deg^2$ of the sky, on a field centred at RA = 10:00:28.6 and DEC = +02:12:21.0 (J2000). The field has been defined by the original coverage in the optical $i_{\rm 814w}$ band with the Advanced Camera for Surveys (ACS) on the {\em Hubble Space Telescope (HST)}. More extensive optical imaging for COSMOS has been taken with broad, narrow and intermediate-band filters for SuprimeCam on the Subaru telescope \citep{tan07}. In the $u^\ast$ band, COSMOS has been observed with Megacam on the Canada-France Hawaii Telescope (CFHT).

In addition, COSMOS has been targeted at mid- and far-IR wavelengths by the {\em Spitzer Space Telescope} \citep{wer04} during the cryogenic mission, as part of the {\em Spitzer} Legacy Program S-COSMOS \citep{san07}. The four-band observations with the Infrared Array Camera \citep[IRAC;][]{faz04} resulted in source catalogues that are $\sim 80\%$ and $\sim 70\%$ complete at mag=23, for $3.6 \, \rm  \mu m$ and $\ffm$, respectively. The observations with the Multiband Photometer for {\em Spitzer} \citep[MIPS;][]{rie04} achieved a depth of $\sim 70 \rm \mu Jy$ at $\tfm$ ($\sim 80\%$ catalogue completeness).

The on-going UltraVISTA survey \citep{mcc12} is providing ultra-deep near-IR images of the COSMOS field in four broad bands ($Y, J, H$ and $K_s$) and a narrow band NB118 \citep{mil13}. The survey strategy is such that it produces alternate deep and ultra-deep stripes oriented N-S, covering a total of $\sim 1.5 \times 1.23 \, \rm deg^2$.   The data used here correspond to the data release version 2 (DR2), which achieves depths of $Ks\approx 24.8$, $H \approx 24.7$, $J \approx 25.1$ and $Y \approx 25.4$ ($5\sigma$; 2~arcsec diameter apertures) on the ultra-deep stripes (Laigle et al., in preparation). These values are 0.7-1.1 magnitudes fainter than the characteristic depths of the UltraVISTA DR1 release discussed by McCracken et al~(2012).

\section{A sample of Spitzer bright  ($[4.5]<23$) sources with UltraVISTA faint ($\ksau$) counterparts in the COSMOS field}
\label{sec-sample}

\subsection{Sample selection and multi-wavelength photometry}
\label{sec-samsel}

We used the publicly available S-COSMOS IRAC catalogue\footnote{http://irsa.ipac.caltech.edu/cgi-bin/Gator/nph-scan?projshort=COSMOS} \citep{ilb10} to select a sample of bright $\ffm$ sources with $[4.5]<23$. To ensure the most reliable source detection and photometry, we only considered sources detected also at $3.6 \, \rm \mu m$, and with an extraction flag equal to 0 in both bands (indicating that the sources are in unmasked areas). 

Independently, we extracted a source catalogue from the UltraVISTA DR2 $K_s$-band mosaic using the software SExtractor \citep{ba96}, imposing a detection threshold of 2.5$\sigma$ over 5 contiguous pixels. Such low threshold is appropriate in this case, as we are only aiming to find counterparts for the robustly detected sources in the IRAC bands.  We measured aperture photometry in 3-arcsec-diameter circular apertures for all the extracted sources in the $K_s$ band, and derived aperture corrections from the curves of flux growth of isolated stars in the field. To perform photometric measurements on all the other UltraVISTA broad bands, we used SExtractor in dual-image mode, using the $K_s$-band mosaic as detection image. The aperture corrections vary between -0.11 and -0.19 mag, depending on the band. 

We preferred using 3-arcsec diameter apertures rather than smaller apertures because a priori we do not know the redshifts of our sources, and too small apertures can be inadequate to recover the photometry of low-$z$ galaxies. Nevertheless, we tested the use of 2-arcsec diameter apertures and found no significant difference in our results for high-$z$ galaxies, which we describe below.  

\begin{figure}
\epsscale{1.1}
\plotone{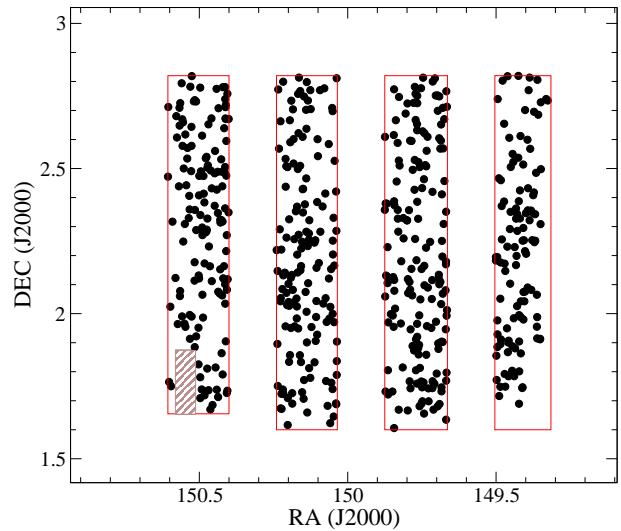}
\caption{Distribution of our $[4.5]<23$ sample with $\ksau$ counterparts in the UltraVISTA/COSMOS field, where the red rectangles indicate the UltraVISTA ultra-deep stripes.  The dashed area at the bottom of the left-hand side stripe indicates a region with no data, due to a faulty detector in the VISTA telescope wide-field camera VIRCAM \citep{sut15}. \label{fig-srcdist}}
\end{figure}

The next step was cross-correlating the {\em Spitzer} IRAC $[4.5]<23$ sources with the UltraVISTA $K_s$-band sources. We limited our study to the UltraVISTA ultra-deep stripes ($\sim 0.8~\rm deg^2$; Figure~\ref{fig-srcdist}) to ensure that we worked with near-IR data of homogeneous depth. The overall identification completeness of our IRAC $[4.5]<23$ sources over these ultra-deep stripes is $\sim 99\%$. 

As the aim here is to study the bright IRAC sources that could not be identified before with the UltraVISTA DR1 dataset, we looked for $[4.5]<23$ sources with a $\ksau$ counterpart. Exclusively for the purpose of identifying sources with faint $K_s$ counterparts, we used a $K_s$ magnitude cut based on the SExtractor `mag\_auto', as this guarantees that we are dealing with a sample complementary to those obtained from the UltraVISTA DR1 images \citep{ilb13}. This selection makes this new sample also complementary to the IRAC galaxy sample studied by Caputi et al.~(2011), in which the identification completeness of the $[4.5]<23$ galaxies was $\sim 96\%$, and less than 1\% of the identified $[4.5]<23$ sources had $\ksau$ at all redshifts, and $<4\%$ at $z\geq 3$. This complementarity is important to assess the corrections that the newly identified bright IRAC galaxies introduce to the GSMF at $z \geq 3$ (cf. \S\ref{sec-gsmf} and \S\ref{sec-gsmfz5}). For all other purposes, we used total magnitudes obtained from aperture magnitudes and corresponding aperture corrections. 

We identified 604 IRAC $[4.5]<23$ sources with  $\ksau$ counterparts within a 1-arcsec matching radius. To avoid dealing with cases with severely contaminated photometry, we excluded from our sample 30 sources that also have a brighter ($K_s^{\rm auto}< 24$) neighbour within 2~arcsec radius. Our final, clean, $[4.5]<23$ and $\ksau$ sample contained 574 sources. Note that, in spite of imposing a matching radius of 1~arcsec, the median separation between the IRAC and $K_s$ band centroids in our sample is only 0.26~arcsec.

We also compiled the CFHT $u^\ast$-band and SUBARU optical ancillary data available for our sources, including the COSMOS broad, narrow and intermediate-band photometry. To obtain this multi-wavelength photometry, we ran SExtractor once again in dual-image mode, using the UltraVISTA $K_s$-band mosaic as detection image, with the same source extraction parameters as before. As for the UltraVISTA data, we measured aperture photometry in 3-arcsec diameter apertures, and derived aperture corrections from the curves of flux growth of isolated stars in the field.

Our finally compiled catalogue contains photometric information for our 574 sources in 30 photometric bands. In total, more than 75\% of our 574  IRAC $[4.5]<23$ sources with $\ksau$ counterparts are detected in the $z$, $z^{++}$ and different optical bands (in addition to being detected in the near- and mid-IR). All the remaining sources are detected in at least six IRAC/UltraVISTA bands. All the magnitudes in our final catalogue are total (obtained from corrected aperture magnitudes) and have been corrected for galactic extinction. As in Ilbert et al.~(2013), we have multiplied all error bars by a factor of 1.5 to account for underestimated photometric errors in SExtractor's output, mainly due to correlations between pixels produced by image re-sampling.

To identify galactic stars, we used a $(B-J)$  versus $(J-[3.6])$ colour-colour diagram, similarly to Caputi et al.~(2011). We found that none of our sources display the characteristic blue $(J-[3.6])$ colours of galactic stars. As we explain in Section \ref{sec-sedz}, we additionally checked for the presence of red dwarf contaminants, but found none either within our sample.

\subsection{Spectral energy distribution (SED) modelling and photometric redshifts}
\label{sec-sedz}

We performed the SED modelling of our sources with a customised $\chi^2$-minimisation fitting code, using Bruzual \& Charlot (2003) synthetic stellar templates for different star formation histories: a single stellar population, and exponentially declining models with characteristic times ranging between $\tau=0.1$ and $5.0 \, \rm Gyr$, all with solar metallicity. For each galaxy, we tested all redshifts between $z=0$ and 7, with a step $dz=0.02$. To account for the galaxy internal extinction, we convolved the stellar templates with the Calzetti et al.~(2000) reddening law, allowing for V-band extinction values $0 \leq A_V \leq 6$. We used the Madau~(1995) prescription to include the effects of the intergalactic medium attenuation at $\lambda_{\rm rest}<1216\, \rm\AA$. As an output of our code, we obtained each galaxy best-fit photometric redshift ($z_{\rm phot}$) and derived parameters, including the stellar mass ($M_{\rm st}$).

In a first step, we performed the SED modelling of our sources over 28 photometric bands, from $u^\ast$ through the IRAC $\ffm$ band. The reason to exclude the longest-wavelength IRAC bands in this step is that, for low-$z$ sources, dust emission (including that from polycyclic aromatic hydrocarbons) can significantly contribute to the photometry beyond $\sim 5 \, \rm \mu m$, so the SED fitting with pure stellar templates may not be adequate. As a priori we did not know the redshifts of our sources, we excluded the longest-wavelength data in our first photometric redshift run. For the SED modelling, when a source was non-detected in a given broad band, we rejected any template that produced a flux above the $3\sigma$ detection limit in that band.  Narrow and intermediate bands were ignored in the cases of non-detections.

We imposed a maximum redshift for each source, according to its detections at short wavelengths, following the criteria explained in Caputi et al.~(2011): for sources detected in the $u*$ band (with $>2 \sigma$ significance), the maximum accepted redshift was $z_{\rm max}=3.6$. For sources significantly detected in $Bj$, $Vj$, and $r^+$, the maximum imposed redshifts were 4.6, 5.6, and 6.4, respectively. We note that, in practice, these maximum redshift constraints had to be applied for a small amount of sources ($\sim2\%$ of the sample), as our code directly produced best-fit redshifts consistent with the short-wavelength photometric detections in the vast majority of cases.

\begin{figure}
\epsscale{1.1}
\plotone{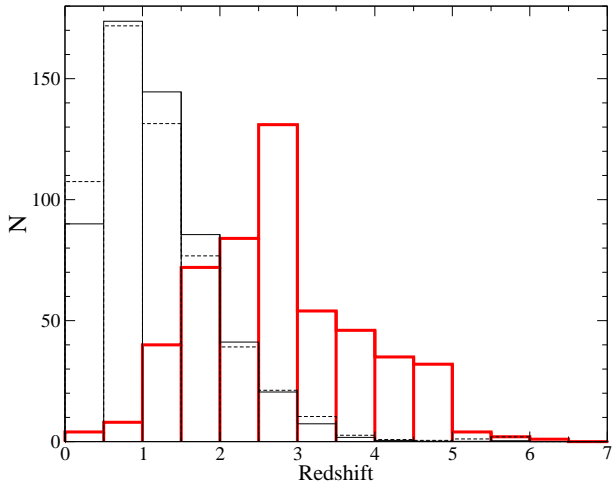}
\caption{Redshift distribution of our $[4.5]<23$, $\ksau$ sample (thick solid histogram). For a comparison, we also show the renormalised redshift distributions of  the $[4.5]<23$, $K_s^{\rm auto}< 24$  galaxies (thin solid histogram), and all $K_s^{\rm auto}< 24$  galaxies (thin dashed histogram). \label{fig-zdist}}
\end{figure}

For all the $z_{\rm phot} \geq 3$ galaxy candidates obtained in the SED modelling run (a total of 252 sources), we performed a second SED modelling considering the 30 photometric bands (i.e., with the full IRAC photometry).  This run with full IRAC photometry was only performed in the cases of reliable IRAC $5.8$ and  $8.0\, \rm \mu m$ detections, i.e., only for magnitudes $<22.5$ and $<22.0$~AB mag, at  $5.8$ and  $8.0\, \rm \mu m$, respectively (which corresponds to $> 3\sigma$ detections). We analysed the SED fitting results for these sources considering the results of the 30-band SED modelling, and also the full $\chi^2$ map in all cases, in order to investigate degeneracies in $A_V-z$ space. 

To accept a $z_{\rm phot} \geq 3$ galaxy candidate, we imposed the criterion that the best-fit solution should be $z_{\rm phot} \geq 3$ both in the 28- and the 30-band SED fitting, when both were available. For the $z_{\rm phot} \geq 5$ candidates, we also requested two additional conditions, namely: i) no secondary $\chi^2$ local minimum should exist at $z_{\rm phot}<5$ within $1\sigma$  of the best-fit solution at $z\geq 5$ (i.e., reduced  $\chi^2$ (secondary $z_{\rm phot}$) - $\chi^2$ (primary $z_{\rm phot}$) $>1$); ii) the median of the marginalized $P(z)$ versus $z$ distribution should also be at $z_{\rm phot} \geq 5$. The latter is especially helpful in cases of broad $P(z)$ distributions, where no secondary $\chi^2$ local minimum is identified within $1\sigma$, but $P(z)$ significantly extends to lower $z$. These are conservative criteria that ensure that we keep only reliable high-$z$ sources in our sample.

Overall, we found that 188 out of 252 $z_{\rm phot} \geq 3$ candidates satisfy all these criteria. For the remaining candidates, we replaced the best-fitting solutions by the 30-band best solutions, when available,  the secondary lower $z_{\rm phot}$  within $1 \, \sigma$ confidence, or the lower median $z_{\rm phot}$ values, with corresponding best-fit parameters.

In the analysis of the $z_{\rm phot} \geq 3$ galaxy candidates with full IRAC photometry, we searched for galaxies with a plausible IR power-law SED component, following the power-law subtraction methodology explained in Caputi~(2013). Such power-law SED component is a clear indication of active galactic nuclei (AGN) at such high $z$, but only the warmest sources at $z\geq 3$ can be identified using data up to $8 \, \rm \mu m$ (the maximum contribution of the AGN-associated IR power law occurs at rest wavelengths $1-2 \, \rm \mu m$, and thus this maximum is shifted beyond observed $\sim 8 \, \rm \mu m$ at $z_{\rm phot} \geq 3$; see discussion in Caputi~2013, 2014). Nevertheless, our aim here is not to make a complete AGN census among our galaxies, but rather to identify cases where the $z_{\rm phot}$ and stellar masses could be affected by a non-stellar component in the SED. So, it is interesting to identify which of our high-$z$ galaxies do manifest an IR  power-law signature.  Among our  $z_{\rm phot} \geq 3$  candidate galaxies, only three display a mid-IR excess that suggests an AGN presence (and one of these has a revised redshift $z_{\rm phot}<3$). 

Recently, it has been pointed out that the presence of emission lines can produce contamination in the selection of high-$z$ galaxies \citep{zac08,sch09}. However, this problem typically arises for galaxies with blue, rather than red, near-/mid-IR colours \citep{deb14}.  We investigated the possible presence of low-redshift interlopers in our $z_{\rm phot} \geq 3$  galaxy sample produced by line emitters by doing an independent photometric redshift run using the public code LePhare \citep{arn99,ilb06}, including emission lines. We found that only two of our $z_{\rm phot} \geq 3$ galaxies have a significantly lower best-fit redshift ($1.25<z<2.6$), while all the rest are confirmed to be at $z_{\rm phot} \geq 3$. We adopted the lower redshifts for the two interlopers found with SED fitting with emission lines.

As a summary, we have 185 galaxies at $z_{\rm phot} \geq 3$, which make $\sim 32\%$ of our total $[4.5]<23$, $\ksau$  sample. Among these high-$z$ sources, we identified 9 reliable galaxy candidates at  $z_{\rm phot} \geq 5$. 

None of our  $z_{\rm phot} \geq 3$ sources appear to be a red dwarf contaminant. We modelled the SEDs of these galaxies using the characteristic stellar templates of M and L stars \citep{ray09}, but in no case these templates produce a good fitting of the observed light. Our $z_{\rm phot} \geq 3$  sample does not contain later-type dwarf contaminants either.  Most T dwarfs and the Y0 dwarfs recently discovered by the {\em Wide-field Infrared Survey Explorer (WISE)} are bright in the mid-IR, but bright in the near-IR as well, and typically have $[3.4]-[4.6] \gsim 1.5$ \citep{eis10,kir11}. Our sources are near-IR faint and all have $[3.6]-[4.5] \lsim 1$.

\begin{figure}
\epsscale{1.1}
\plotone{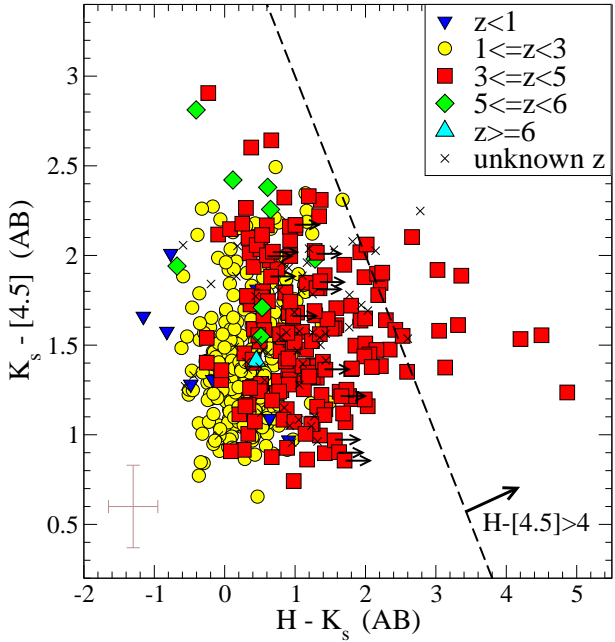}
\caption{$K_s-[4.5]$ versus $H-K_s$ colour-colour diagram for our $[4.5]<23$, $\ksau$ sources at different redshifts. The dashed line delimits the region corresponding to  $H-[4.5]>4$ colours, which is the colour-cut imposed for the sample selection in Caputi et al.~(2012). Median error bars on the colours are shown in the bottom left corner of the plot. \label{fig-colz}}
\end{figure}

\subsection{Redshift distribution and IR colours}

Figure \ref{fig-zdist} shows the resulting redshift distribution of our $[4.5]<23$, $\ksau$ sample. Virtually all our galaxies lie at redshifts  $z_{\rm phot}> 1$, with a clear peak at redshifts  $2.5 < z_{\rm phot} <3.0$. This bias towards high redshifts is a consequence of the red colours imposed by our double magnitude-limited selection.  This effect is similar to that observed in classical `extremely red galaxies', whose redshifts typically are $z \gsim 1.0$ \citep{cap04,geo06}, and dust-obscured galaxies (DOGs), which mainly lie at $z \gsim 2$ \citep{dey08,pop08}.

In addition, the redshift distribution has a significant tail at $z_{\rm phot} \geq 3$, containing $\sim 32\%$ of the galaxies in our sample. In this high-$z$ tail, there are 9 galaxies at redshifts $z_{\rm phot} \geq 5.0$. 

About 10\% of our galaxies have no $z_{\rm phot}$ determination. There are two reasons for this: for 2\% of our sample, the minimum $\chi^2$ value is too high to trust the resulting $z_{\rm phot}$. For the other 8\% of sources, the resulting probability distribution in redshift space $P(z)$ is basically flat at $z\gsim 2$, so it is not possible to decide on any redshift.

As our galaxies are faint at optical and near-IR wavelengths, it is currently very difficult to obtain spectroscopic confirmations for our redshifts. We have cross-correlated our catalogue with a compilation of all available sources with spectroscopic redshifts in COSMOS, including those in the zCOSMOS \citep{lil07} and VUDS surveys \citep{lef15}, and data taken with the DEIMOS and FMOS spectrographs (Capak et al., in preparation; Kartaltepe et al., in preparation), among others. We found a total of six matches, and in four of these cases our photometric redshifts are in very good agreement with the spectroscopic redshifts within the error bars (these redshifts range between $z=2.10$ and $z=3.88$). The other two sources have low signal-to-noise spectra, with (only tentative) redshifts measurements which do not agree with our photometric determinations (which are $z_{\rm phot}=1.68$ and 4.08). As the spectral quality of these two sources is not really useful for a diagnostic, we considered that our photometric redshift estimates are likely correct.

In Figure \ref{fig-colz}, we show  a $K_s-[4.5]$ versus $H-K_s$ colour-colour plot, where we have segregated our galaxies by redshift. We see that our galaxies span a wide range of $K_s-[4.5]$ colours, including a minority with $K_s-[4.5] \leq 1$. This is because our colours are based on total magnitudes (obtained from aperture magnitudes, plus corrections), while the $\ksau$ cut applied in our sample selection refers to the SExtractor $K_s$ `mag\_auto'. (The median difference between the $K_s$ total and the $K_s^{\rm auto}$ magnitudes in our sample is about $-0.07$~mag). 

The $K_s-[4.5]$ colour, in particular, is only mildly related to the redshift for the $[4.5]<23$ galaxies. About two-thirds of the sources with $K_s-[4.5]>2$ lie at $z_{\rm phot} \geq 3$, and the other third are mostly $2 < z_{\rm phot}< 3$ galaxies with extinction values $A_V\approx 3.0$. The $H-K_s$ colour is potentially a better redshift discriminator: at least two-thirds of the $H-K_s \gsim 1$ sources lie at $3 \leq z_{\rm phot} <5$, and only 13\% are confirmed at lower redshifts (there are about 20\% of sources with these colours that have no redshift estimates in our sample). The $z_{\rm phot} \geq 5$ galaxy candidates, instead, are characterised by flat near-/mid-IR spectra, with $H-K_s$ colours around 0, similarly to most $z_{\rm phot}<3$ galaxies. 

Only a small fraction of our sources display the extremely red $H-[4.5]>4$ colours characterising the galaxies studied by Caputi et al.~(2012). This is mainly a consequence of the insufficient depth of the UltraVISTA DR2 images to probe such red sources. The bulk of these extremely red galaxies are expected to be among the $\sim$1\% of $[4.5]<23$ sources that remain unidentified in the UltraVISTA DR2. The sources that do satisfy $H-[4.5]>4$ in our current sample appear to be at $3 \leq z < 5$, consistently with the results of Caputi et al.~(2012).

The majority of the $\sim10$\% of sources with no redshift estimate in our sample display similarly red $H-K_s$ colours as the $3 \leq z <5$ sources, so they also likely lie at these redshifts. Only a few sources with no redshift estimates display similar colours to the $z=5-6$ candidates, but in this case we cannot conclude on a plausible high $z$, as many sources at $z<3$, and a minority at $3 \leq z <5$, also have similar colours.

\section{Massive galaxies at $3 \leq \lowercase{z} < 5$}
\label{sec-massz35}

\subsection{Properties and comparison with near-IR brighter, massive galaxies at similar redshifts}

\begin{figure}
\epsscale{1.1}
\plotone{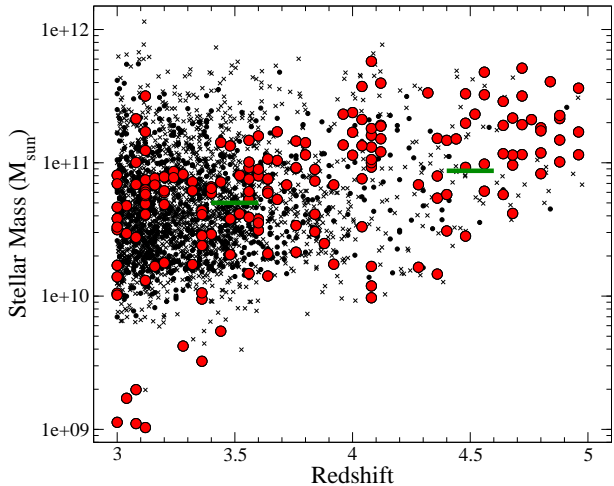}
\caption{Estimated stellar masses versus photometric redshifts for our $[4.5]<23$, $\ksau$ galaxies at $3 \leq z < 5$ (red filled circles). The data points in the background indicate $[4.5]<23$, $K_s^{\rm auto}< 24$ galaxies from two different samples: the UDS sample analysed by Caputi et al.~(2011) (small black dots) and the UltraVISTA DR1 sample analysed by Ilbert et al.~(2013) (crosses). The stellar masses derived by Caputi et al.~(2011) have been multiplied by a factor of 1.24 to obtain a crude conversion from BC07 to BC03 templates, while the Ilbert et al. (2013) stellar masses have been multiplied by a factor of 1.7 to convert from a Chabrier to a Salpeter IMF over $(0.1-100) \,\rm M_\odot$. The horizontal lines indicate the stellar mass completeness limits imposed by the $[4.5]<23$ magnitude cut at $z\sim3.5$ and 4.5.
 \label{fig-stmvsz}}
\end{figure}

A total of 176 galaxies in our $[4.5]<23$, $\ksau$ sample lie at redshifts $3 \leq z <5$. Figure~\ref{fig-stmvsz} shows the derived stellar masses versus photometric redshifts for all these galaxies. We see that the stellar masses range from $\sim 10^9$ to $\sim 6 \times 10^{11} \, \rm M_\odot$, and about 66\% of our galaxies are quite massive, with $M_{\rm st}>5 \times 10^{10} \, \rm M_\sun$.  At redshifts $z>4$, almost $90\%$ of our galaxies display such large stellar masses. The horizontal lines centred at $z=3.5$ and 4.5 in Fig.~\ref{fig-stmvsz} indicate the stellar mass completeness limits imposed by the $[4.5]<23$ magnitude cut at these redshifts. These limiting stellar masses correspond to galaxies described by a single-stellar-population template, with no dust, and with an age equal to the age of the Universe at those redshifts.

For a comparison, in the same plot we show the stellar masses obtained for $[4.5]<23$ galaxies with brighter $K_s$ counterparts in COSMOS and the Ultra Deep Survey (UDS) field. By simple inspection, it becomes clear that the new galaxies analysed here contribute significantly to the population of massive galaxies at $z \geq 3$. This is especially true at $4 \leq z<5$, where our newly analysed $[4.5]<23$, $\ksau$ galaxies increase by a factor $\gsim 2$ the number density of known massive galaxies. In the next Section, we investigate the effects that this substantial increment has on the GSMF at these redshifts.

The maximum extinction value associated with the best-fit solutions of our $3 \leq z < 5$ sources is $A_V=2.7$, and the median value is of only $A_V=1.2$. Thus, we conclude that the extinction values keep moderate even for the reddest sources in our sample (with respect to the maximum $A_V$ value allowed in our SED fitting grid, i.e., $A_V=6$). 

It is interesting to compare these extinctions and other average SED properties with those obtained previously for massive galaxies  in other sample selections at $3 \leq z <5$.  For the extremely red sources with $3 \leq z < 5$   in the Caputi et al.~(2012) sample, the median derived extinction was $A_V=2.2$, which is significantly higher than the median value found for our sample here, but still relatively moderate. In addition, the $3 \leq z <5$ galaxies in Caputi et al.~(2012) had associated best-fit ages $\gsim 0.5  \, \rm Gyr$  in all cases, while around a quarter of the galaxies in our current $3 \leq z < 5$ sample have best-fit younger ages. On the other hand, in the case of the $[4.5]<23$ galaxies at $3 \leq z <5$ included in the Caputi et al.~(2011) sample, which had no colour restriction, the median extinction was $A_V=0.8$, and 60\% of the sample had best-fit ages younger than $\sim 0.5  \, \rm Gyr$.

Thus, we conclude that the colours of massive galaxies at $3 \leq z < 5$ are the result of a combination of age and dust extinction: redder galaxies are typically both older and dustier than bluer ones.

\subsection{The GSMF at $3 \leq z < 5$}
\label{sec-gsmf}

\subsubsection{An updated calculation of the GSMF high-mass end at $3 \leq z < 5$}

\begin{figure*}
\epsscale{1.1}
\plotone{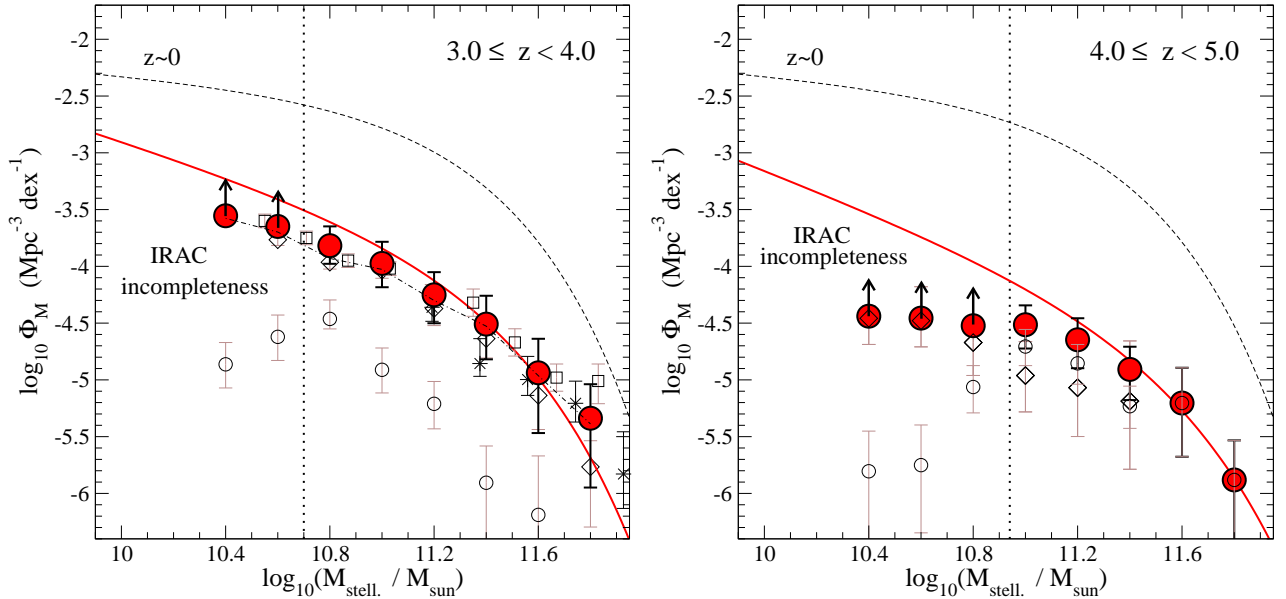}
\caption{GSMF at $3.0 \leq z < 4.0$ ({\em left}) and $4.0 \leq z < 5.0$ ({\em right}), obtained by considering the results of previous $K_s^{\rm auto}< 24$ surveys and the new $[4.5]<23$, $\ksau$ galaxies analysed here.  The large filled circles correspond to the total GSMF, while the small open circles show the contribution of only our $[4.5]<23$, $\ksau$ galaxies, as computed with the $V/V_{\rm max}$ method. Other symbols correspond to previous $K_s^{\rm auto}< 24$ surveys: diamonds (Caputi et al.~2011) and squares (Ilbert et al.~2013), with the dashed-dotted line on the left-hand side panel indicating the average between the two determinations at $3.0 \leq z < 4.0$.  The GSMF determination by Muzzin et al.~(2013; asterisks) is also shown for a comparison. All $V/V_{\rm max}$ data points correspond to a Salpeter IMF over $(0.1-100) \,\rm M_\odot$, and the stellar masses from Caputi et al.~(2011) have been multiplied by a factor of 1.24 to obtain a crude conversion from BC07 to BC03 templates. The solid red line in each panel indicates the resulting GSMF obtained with the STY maximum likelihood analysis, assuming a single Schechter function. This maximum likelihood analysis has been performed on a combination of Caputi et al.~(2011) sample and the new galaxies analysed in this paper. The vertical dotted line in each panel indicates the stellar mass completeness limit imposed by the IRAC $[4.5]=23$ magnitude cut. The dashed line indicates the local GSMF \citep{col01}.
\label{fig-gsmf35}}
\end{figure*}

A main goal of this work is to assess the importance of the newly identified $3 \leq z < 5$ galaxies with respect to the population of massive galaxies known at these redshifts. For this, we analyse their contribution to the high-mass end of the GSMF. We note that for this analysis we did not exclude any source as a result of the presence of an IR AGN power-law component. We rather considered corrected stellar masses, computed after power-law subtraction from the SED (see Caputi~2013), in the few necessary cases discussed before.

Figure \ref{fig-gsmf35} shows the GSMF at $3 \leq z <4$ and $4 \leq z <5$. The red circles in  Figure \ref{fig-gsmf35} and values in Table \ref{tab-vmaxz35} indicate the updated  $V/V_{\rm max}$ GSMF values, obtained by adding the contribution of our $[4.5]<23$, $\ksau$ galaxies to the average GSMF values obtained by other recent analyses of large-area, near-/mid-IR surveys. These previous GSMF determinations are based on the IRAC survey of the UDS field  \citep{cap11} and UltraVISTA DR1 in COSMOS \citep{ilb13}.  From all these studies, the analysed sources are complementary to our current sample (i.e., they have $K_s^{\rm auto}< 24$), and the differently obtained GSMF are in very good agreement amongst themselves. In the case of the UDS data, we have re-computed the GSMF values in the redshift bins considered here after excluding a minor fraction of sources with $K_s^{\rm auto}> 24$ in the Caputi et al.~(2011) sample.

For a correct joint analysis and comparison, we have converted the UltraVISTA previous determinations to the same IMF used here, i.e., Salpeter over stellar masses $(0.1-100) \, \rm M_\odot$, and the stellar mass values from Caputi et al. (2011) have been multiplied by a factor of 1.24 to provide a crude correction from the 2007  to the 2003 version of the Bruzual \& Charlot templates. To obtain the average GSMF of these previous surveys, we have linearly interpolated the UltraVISTA DR1 values at the stellar-mass bin centres considered here. 

To consider the $V/V_{\rm max}$  contribution of each of our new galaxies, we took into account the fact that they are selected with a magnitude limit $[4.5]<23$. This criterion imposes a maximum redshift at which the galaxy would be in the sample. On the other hand, the $\ksau$ criterion imposes a {\em minimum} redshift at which the galaxy would be included. However, since here we are analysing jointly previously selected $K_s^{\rm auto}< 24$ galaxy samples and our new sample with $\ksau$, then this extra correction is not necessary.

The red solid lines in Figure \ref{fig-gsmf35} and parameter values in Table \ref{tab-schech35} indicate the result of an independent computation of the GMSF, obtained by applying the STY \citep{sty79} maximum likelihood analysis, and assuming that the GSMF has the shape of a single Schechter function \citep{sch76}. For this GSMF computation, we considered the complementary IRAC galaxy sample in the UDS from Caputi et al.~(2011), along with the new UltraVISTA COSMOS sample analysed here (each with their corresponding magnitude limits and surveyed areas). The STY method is a parametric technique that, in contrast to the $V/V_{\rm max}$ method, involves no data binning and has no implicit assumption on a uniform galaxy spatial distribution. However, it has the disadvantage that it does not automatically provide the normalisation of the GSMF, which rather has to be provided as an input parameter.  Here, we considered the number density of galaxies with $M_{\rm st}>10^{11} \, \rm M_\odot$ in our sample at different redshifts to compute this normalisation. It is worth reminding the reader that the curves resulting from the STY method do not constitute a fitting to the $V_{\rm max}$ points. We refer the reader to Caputi et al.~(2011) for more details about the GSMF computation using the STY method. 

\begin{deluxetable}{ccc}
\tabletypesize{\scriptsize}
\tablecaption{GSMF values obtained with the $V/V_{\rm max}$ method at $3 \leq z < 5$.  \label{tab-vmaxz35}}
\tablehead{\colhead{$\log_{10} \rm (M/M_\odot)$} &  \multicolumn{2}{c}{$\log_{10} (\rm \Phi_M / Mpc^{-3} \, dex^{-1})$} \\
  &  $3.0 \leq z <4.0$ & $4.0 \leq z < 5.0$
}  
\startdata
10.40 & $>-3.56$  & $>-4.44$ \\
10.60 & $>-3.65$ &  $>-4.46$  \\
10.80 & $-3.82^{+0.17}_{-0.16}$ &  $>-4.52$  \\
11.00 & $-3.97^{+0.19}_{-0.21}$ &  $-4.51^{+0.17}_{-0.21}$   \\  
11.20 & $-4.25^{+0.20}_{-0.25}$ &  $-4.65^{+0.19}_{-0.25}$   \\ 
11.40 & $-4.51^{+0.25}_{-0.31}$ &  $-4.91^{+0.20}_{-0.27}$   \\ 
11.60 & $-4.94^{+0.30}_{-0.53}$ &  $-5.20^{+0.31}_{-0.47}$  \\
11.80 & $-5.34^{+0.30}_{-0.61}$ &  $-5.88^{+0.35}_{-0.70}$ 
\enddata
\end{deluxetable}

\begin{deluxetable*}{lccccc}
\tabletypesize{\scriptsize}
\tablecaption{Schechter function free parameter values obtained for the GSMF computed with the maximum likelihood STY analysis.   \label{tab-schech35}}
\tablehead{\colhead{Redshift} & \colhead{$\alpha$} & \colhead{$\rm M^\ast (M_\odot)$} &  \colhead{$\rm \Phi^\ast_M  (Mpc^{-3} \, dex^{-1})$} & \colhead{$\rm \rho_M^{\rm compl.} (M_\odot Mpc^{-3})$\tablenotemark{b}}  & \colhead{$\rm \rho_M^{\rm TOTAL} (M_\odot Mpc^{-3})$\tablenotemark{c}}  
}
\startdata
$3.0 \leq z < 4.0$ & $1.72 \pm 0.06$  & $(1.82^{+0.18}_{-0.16}) \times 10^{11}$ &  $(1.62^{+0.85}_{-0.48}) \times 10^{-4}$ & $(1.11^{+0.58}_{-0.33}) \times 10^7 $ &  $(3.56^{+1.87}_{-1.06}) \times 10^7 $   \\
\tablenotemark{a} & $1.58 \pm 0.06$  & $(1.79^{+0.16}_{-0.21}) \times 10^{11}$ &   $(1.23^{+0.95}_{-0.26}) \times 10^{-4}$ &  $(7.86^{+6.03}_{-1.65}) \times 10^6 $ & $(1.92^{+1.47}_{-0.40}) \times 10^7$\\
$4.0 \leq z < 5.0$ & $1.88^{+0.12}_{-0.18}$ & $(2.40^{+0.48}_{-0.74}) \times 10^{11}$ &  $(4.39^{+2.49}_{-1.09}) \times 10^{-5}$   & $(3.41^{+1.93}_{-0.84}) \times 10^6 $ &  $(2.10^{+1.19}_{-0.52}) \times 10^7$ \\
\tablenotemark{a} & $1.58^{+0.12}_{-0.16}$ & $(2.00^{+1.16}_{-0.59}) \times 10^{11}$ &  $(6.14^{+3.55}_{-1.52}) \times 10^{-5}$   & $(3.32^{+1.92}_{-0.81}) \times 10^6 $ &  $(1.07^{+0.62}_{-0.26}) \times 10^7$
\enddata
\tablenotetext{a}{The second row for each redshift bin provides the values obtained after correction for Eddington bias.}
\tablenotetext{b}{Stellar mass density values obtained by integrating the resulting Schechter functions above stellar mass completeness.}
\tablenotetext{c}{Stellar mass density values obtained by integrating the resulting Schechter functions for stellar masses $M_{\rm st}>10^8 \, \rm M_\odot$.}
\end{deluxetable*}

Table \ref{tab-schech35} contains the Schechter function parameter values obtained with the maximum likelihood STY analysis. In this analysis, we have left both the $\alpha$ and $M^\ast$ parameters free.  The values of $\rm \Phi^\ast_M$ and $\rm \rho_M$ carry error bars that include the largest errors among the $1\sigma$ errors of the maximum likelihood analysis and the mock realizations described in Section~\ref{sec-edd}, and a fixed 20\% fiducial error to account for cosmic variance (following the determinations of Ilbert et al.~(2013) for massive galaxies).

\begin{figure}
\epsscale{1.1}
\plotone{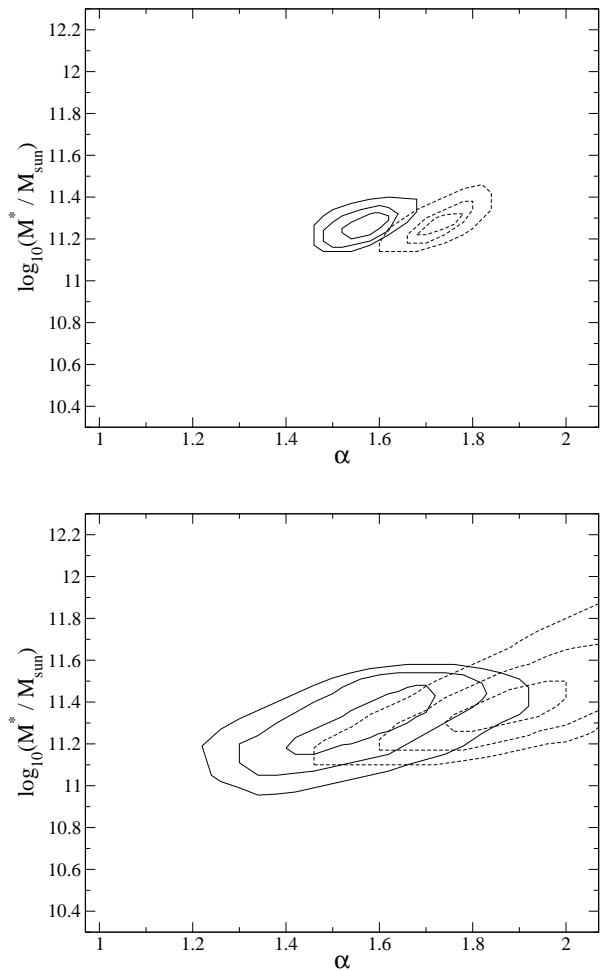}
\caption{ \label{fig-lkh} The 1$\sigma$, 2$\sigma$ and 3$\sigma$ confidence levels for the maximum likelihood free parameters at $3.0 \leq z < 4.0$ ({\em top}) and $4.0 \leq z < 5.0$ ({\em bottom}). The dashed curves correspond to the original maximum likelihood analysis performed considering a Schechter function, i.e., with no correction for Eddington bias.  The solid curves correspond to the Eddington-bias-corrected maximum likelihood analysis, i.e., taking into account the convolution of the Schechter function with Gaussian kernels. }
\end{figure}

We show the 1$\sigma$, 2$\sigma$ and 3$\sigma$ confidence levels on the $\alpha-M^\ast$ plane in Figure~\ref{fig-lkh}. There is the misconception that $\alpha$, commonly referred to as the faint-end slope, cannot be left as a free parameter unless the low-mass end of the GSMF can be well constrained. However, this is not the case, for two reasons: first, the value of $\alpha$  affects exclusively the GSMF low-mass end only when it is close to 1, as the Schechter function is proportional to $(M/M^\ast)^{(1-\alpha)} \times \exp(-M/M^\ast)$. In this case, the effect on the high-mass end is basically negligible. Instead, when $\alpha$ is sufficiently larger than 1, it has two effects: it governs the low-mass end slope and also modulates the exponential decline in the high-mass end. 

Second, the STY method takes into account the survey limiting magnitude, providing a suitable extrapolation to stellar masses below completeness when the regime around and below $M^\ast$ is reasonably well sampled (see Fig. \ref{fig-gsmf35}). This does not happen when a Schechter function is used to do a simple fitting of the $V/V_{\rm max}$ data points, which is the most typical case in the literature.  Of course, having a galaxy sample that represents significantly both the low- and high-mass end would be the ideal situation for computing the Schechter function parameter values.  Sampling a wide range of stellar masses is also necessary to probe, for instance, whether the GSMF follows a single or double Schechter function, but this is beyond the scope of this paper. In spite of not reaching the low-stellar-mass end,  our sample stellar mass completeness limits are below $M^\ast$ at $3 \leq z <5$, so the computation of the Schechter $\alpha$ parameter is meaningful.  The agreement of our maximum likelihood analysis results with previous literature values suggests that this is indeed the case.

\begin{figure}
\epsscale{1.1}
\plotone{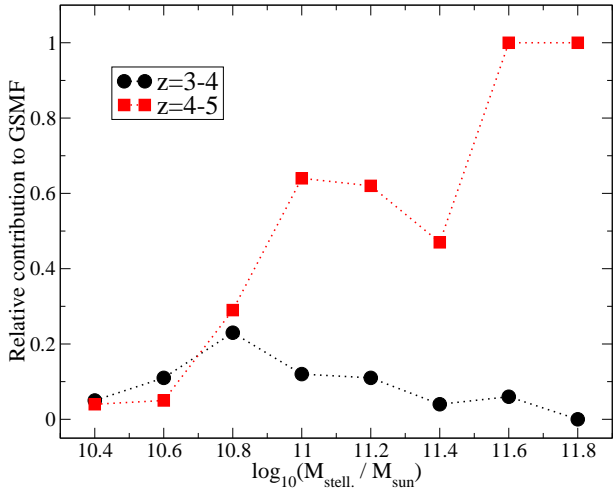}
\caption{Relative contribution of our new $[4.5]<23$, $\ksau$ galaxies to the total GSMF at different stellar masses, at both $3.0 \leq z < 4.0$ and $4.0 \leq z < 5.0$. At $3.0 \leq z < 4.0$, the contribution is small at all stellar masses, while at  $4.0 \leq z < 5.0$ it becomes important at stellar masses above $\sim 6 \times 10^{10} \, \rm M_\odot$. \label{fig-vmaxfr}}
\end{figure}

In Figure~\ref{fig-vmaxfr}, we show the relative contribution of our $[4.5]<23$, $\ksau$ galaxies to the total GSMF. At $3 \leq z <4$,  we see that our new galaxies produce a basically negligible correction to the high-mass end of the GSMF, indicating that a $K_s^{\rm auto}< 24$ survey contains the vast majority of massive galaxies at these redshifts. In fact, as we discuss below, deeper near-IR surveys appear to have a relatively small additional contribution even at intermediate stellar masses  ($10^{10} \lsim M_{\rm st} \lsim 10^{11} \, \rm M_\odot$) at these redshifts.

Instead, the contribution of the $[4.5]<23$, $\ksau$ galaxies to the GSMF high-mass end at $4 \leq z <5$ appears to be more significant. Notably, it allows us to constrain the GSMF at very high stellar masses ($M_{\rm st} \gsim 3 \times 10^{11} \, \rm M_\odot$), which was not possible with the sample of IRAC galaxies with brighter $K_s$ counterparts in Caputi et al.~(2011). 
We have only 11 galaxies with such high stellar masses within our sample at $4 \leq z <5$ in the  UltraVISTA ultra-deep stripes, which indicates that these galaxies are indeed rare. Overall, our new galaxies account for $\gsim 50\%$ of the galaxies with stellar masses $M_{\rm st} \gsim 6 \times 10^{10} \, \rm M_\odot$ at $4 \leq z <5$. Therefore, we conclude that finding the bulk of massive galaxies at $z \geq 4$ requires ultra-deep near-IR surveys covering large areas of the sky.

\begin{figure}
\epsscale{1.1}
\plotone{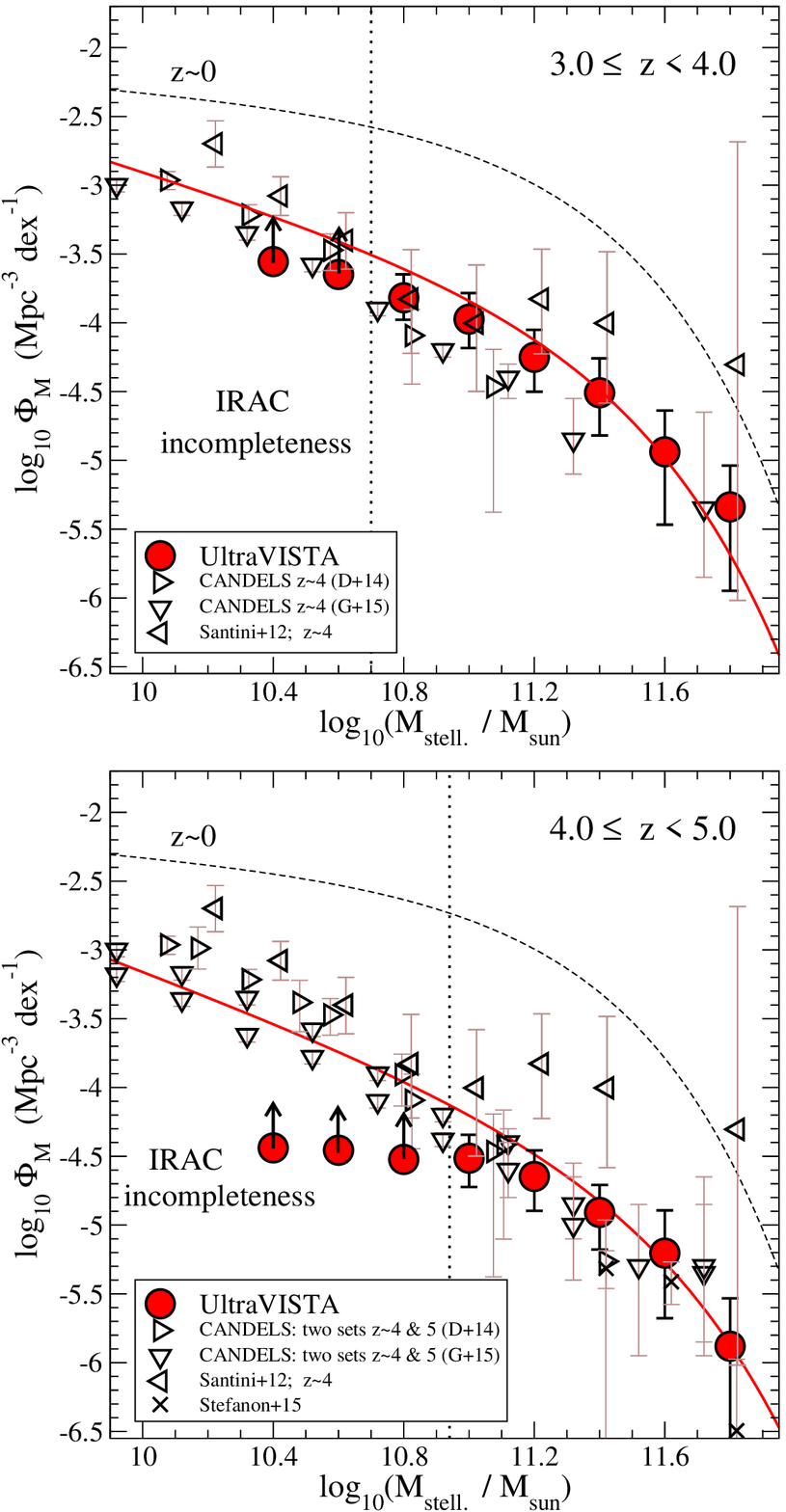}
\caption{GSMF obtained in this paper at $3.0 \leq z < 4.0$ and $4.0 \leq z < 5.0$, compared to those obtained with datasets from the CANDELS survey \citep{dun14,gra15} and other recent works \citep{san12,ste15}. No GSMF determination is available for CANDELS and Stefanon et al.~(2015) at $z<4$, so we only show their results at $z \geq 4$. As in Fig.~\ref{fig-gsmf35}, the vertical dotted line in each panel indicates the stellar mass completeness limit imposed by our IRAC magnitude cut. Stellar masses correspond to a Salpeter IMF over $(0.1-100) \,\rm M_\odot$ in all cases.  \label{fig-gsmfcomp35}}
\end{figure}

Our present STY analysis based on the combination of the UDS IRAC galaxy sample and our current UltraVISTA  $[4.5]<23$, $\ksau$ galaxies yields: $\alpha=1.72 \pm 0.06$ at $3 \leq z < 4$, and $\alpha=1.88^{+0.12}_{-0.18}$ at $4 \leq z < 5$, confirming that $\alpha$ appears to be significantly higher at high $z$ than at $z=0$, as it was found in several previous studies \citep[e.g.][]{kaj09,cap11,san12}. The $\alpha$ value that we obtained at $3 \leq z < 4$ is slightly lower (in absolute value) than that obtained in Caputi et al.~(2011). This is mainly because we have improved the statistics here by considering  broader redshift bins (in Caputi et al.~2011 we used three different redshift bins, i.e., $3.0 \leq z <3.5$, $3.5 \leq z < 4.25$, and $4.25 \leq z < 5.0$, for the GSMF computation). At $4 \leq z<5$, our obtained $\alpha$ value ($\alpha=1.88^{+0.12}_{-0.18}$) is very similar to that obtained by Caputi et al.~(2011) at similar redshifts, even when the new galaxies make a significant contribution to the GSMF at these redshifts. We do obtain a higher $M^\ast$ value than Caputi et al.~(2011), although still consistent within the error bars.

Caputi et al.~(2011) found that a single power-law shape and a Schechter function could not be differentiated as a functional form for the GSMF at $4.25 \leq z <5.0$ in the maximum likelihood STY analysis. And they argued that this could be due to the insufficient sampling of the GSMF at these redshifts. In our present analysis of the GSMF at $4.0 \leq z <5.0$, a single power-law shape is discarded with $> 5 \sigma$ confidence with respect to the Schechter function, indicating that the latter functional form produces a better representation of the GSMF shape up to at least $z=5$.

\subsubsection{Comparison with other recent studies}

It is interesting to compare our new results for the GSMF at $3.0 \leq z <5.0$ with the latest determinations found in the literature. Figure~\ref{fig-gsmfcomp35} shows our obtained GSMF along with that obtained using ultra-deep near-IR datasets from the CANDELS survey \citep{dun14,gra15}, and other recent works \citep{san12,ste15}. We show the computed $V/V_{\rm max}$ points in the different cases,  as this is the clearest and most direct way of comparing two GSMF determinations. 

We see that our GSMF $V/V_{\rm max}$ values above our stellar-mass completeness thresholds are in good agreement with the values obtained with CANDELS at similar redshifts, and also the Stefanon et al.~(2015) values, within the error bars. Note that, in the upper panel of Figure~\ref{fig-gsmfcomp35}, our  $V/V_{\rm max}$ points are systematically higher than the CANDELS values above our stellar mass completeness limit because the mean redshifts of the samples are slightly different ($z\sim 3.5$ and 4, respectively). 

In contrast, at high stellar masses, the $V/V_{\rm max}$ points of Santini et al.~(2012) are systematically higher than all other determinations (although still consistent within the error bars at $3 \leq z <4$). Their GSMF has been computed analysing a pencil-beam survey on the Chandra Deep Field South Early Science Release (ERS) area, which is too small to study the GSMF high-mass end. At $M_{\rm st}<10^{11} \, \rm M_\odot$,  Santini et al. $V/V_{\rm max}$ points appear to be in better agreement with the other results, including our own.

The comparison of the individual Schechter function free parameters is also a common practice in the literature, but one should always keep in mind that $\alpha$ and $M^\ast$ are coupled, so the discussion of each of them separately has to be taken with care. Often, a comparison of these parameters separately is misleading, resulting in wrong conclusions about apparent discrepancies between different studies. With this caveat in mind, we compare the $\alpha$ values derived here with those obtained in CANDELS.

We obtained $\alpha = 1.72\pm0.06$ at $z\sim 3.5$, and $\alpha = 1.88^{+0.12}_{-0.18}$ at $z\sim 4.5$. These values are consistent with those derived by Duncan et al.~(2014) and Grazian et al.~(2015), within the error bars. At $z\sim4$,  Duncan et al. obtained $\alpha=1.89^{+0.15}_{-0.13}$, while Grazian et al. derived $\alpha=1.63\pm0.05$. At $z\sim 5$, the values derived by these authors are $1.74^{+0.41}_{-0.29}$ and $1.63\pm0.05$, respectively. Note that Grazian et al. quoted values are those obtained after correction for Eddington bias, which explains why they are lower than the others. As we explain in Section \ref{sec-edd}, our own $\alpha$ values also become similarly lower after considering this correction.

Therefore, all these results taken together confirm the now well-known fact that the derived $\alpha$  (absolute) values are significantly larger at high redshifts than in the local Universe. The dispersion between the mean $\alpha$ values derived from different datasets is still at the $\sim 10-20\%$ level. This is due to the selection effects intrinsic to the different analysed samples, and also the different ways of deriving the Schechter function characteristic parameter values.  Here we derive these values through an STY maximum likelihood analysis, while Duncan et al.~(2014) and Grazian et al.~(2015) simply fitted the $V/V_{\rm max}$  points. In the latter, the highest stellar mass bins, with large error bars, are basically ignored, while the STY method provides a fairer weight for galaxies of different stellar masses.

\begin{figure}
\epsscale{1.1}
\plotone{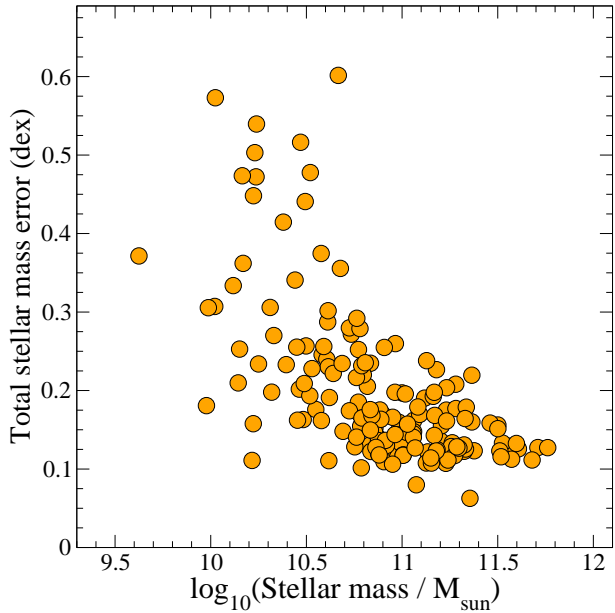}
\caption{Total stellar mass errors versus stellar mass for the $3.0 \leq z < 5.0$ galaxies in our sample. These errors are the sum in quadrature of the stellar mass errors produced by the $z_{\rm phot}$ uncertainties and the errors at fixed $z_{\rm phot}$. In most cases, the total errors are dominated by the component due to $z_{\rm phot}$ uncertainties. The stellar mass errors at fixed redshift only have a contribution of $\sim 0.06$~dex, while most total values are larger than this number.
\label{fig-stmerr}}
\end{figure}

\subsubsection{Error analysis and the Eddington bias in the GSMF determination}
\label{sec-edd}

The errors in the photometric redshifts and stellar masses of our galaxies propagate into uncertainties in the GSMF determination. To analyse this effect, we repeated the STY maximum likelihood 100 times on different mock galaxy catalogues. Each of these mock catalogues contains the `same' galaxies as the original catalogue, but with randomly generated redshifts and stellar masses within the error bars.   To assign a photometric redshift for each galaxy in the mock catalogue, we computed a random value following a Gaussian distribution around the real, best-fit photometric redshift, with an r.m.s. given by the $1\sigma$ confidence interval of the $P(z)$ distribution resulting from each galaxy SED fitting. This treatment is adequate for red sources, as the typical $P(z)$ distribution is broad around a single best-redshift peak, rather than having multiple peaks with similar probability. 

The $z_{\rm phot}$ errors automatically produce variations on the derived stellar masses, which we recomputed consistently at each new mock redshift. Hence, we considered that the $\pm 1\sigma$ error of each stellar mass, due to the redshift variations, was given by the values encompassing 68\% of the resulting stellar masses for the mock $z_{\rm phot}$ of each galaxy, around the `real' stellar mass value. 

In addition to this, we considered that the stellar mass of each galaxy was prone to an additional error, at fixed redshift, given by a Gaussian distribution with a fixed width of $0.30 \times M_{\rm st}$, i.e., $\sigma=0.15 \times M_{\rm st}$.  This is the maximum error on $M_{\rm st}$ observed for $\gsim 68\%$ of our galaxies at fixed redshift, similarly to the findings in the Caputi et al.~(2011) sample. In fact, this additional error constitutes a relatively minor contribution to the total stellar mass error budget, which is dominated by the errors produced by the photometric redshift uncertainties in most cases, as can be seen in Fig.~\ref{fig-stmerr}.

\begin{figure}
\epsscale{1.1}
\plotone{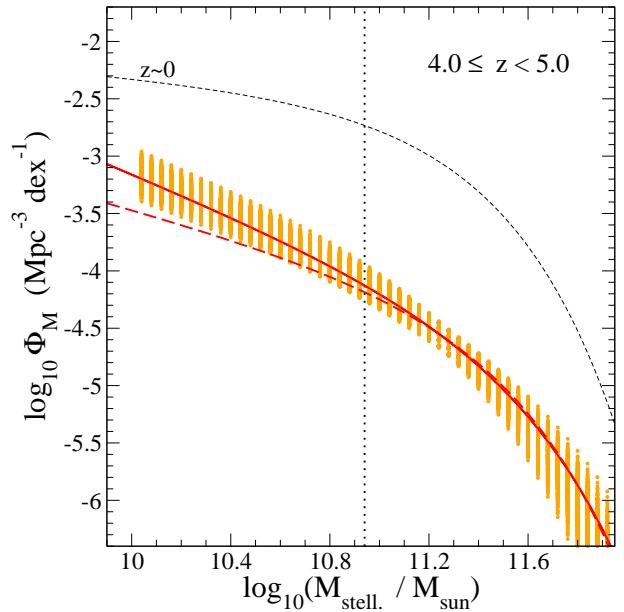}
\caption{Full range of GSMF at $4.0 \leq z < 5.0$ obtained from 100 mock galaxy catalogues (filled orange dots). The red solid line indicates the GSMF obtained applying the STY maximum likelihood analysis on our real sample. This plot reflects the uncertainties on the GSMF produced by the $z_{\rm phot}$ and stellar mass errors. The long-dashed line corresponds to the Eddington-bias corrected GSMF.
\label{fig-gsmfedd45}}
\end{figure}

At $3 \leq z <4$, we found that the resulting scatter in the  GSMF determination is very small. Instead, at $4 \leq z < 5$, the scatter observed in the GSMF is more important (Fig.~\ref{fig-gsmfedd45}). The $\alpha$ values obtained in the different realizations of the GSMF vary between $\alpha=1.64$ and 2.02. We adjusted the error bars of our $\alpha$ value quoted in Table~\ref{tab-schech35} to account for this possible range (which is larger than the range given by the 1$\sigma$ errors of the maximum likelihood fitting on the real catalogues).

On the other hand, the GSMF determination can potentially be affected by the so-called Eddington bias \citep{edd13}. This is a consequence of the errors in the photometric redshifts and stellar mass estimates, which introduce non-negligible scatter in the GSMF. This effect becomes significant when the errors are very large, and/or the number of galaxies considered to compute the GSMF is low. Because of the latter, the Eddington bias affects mostly the GSMF high and low-mass ends, typically producing a flattening which may result in an artificially high (absolute) $\alpha$ value.  
 
To investigate the effect of the Eddington bias, we considered that the observed function describing the GSMF is in fact the convolution of the `real' GSMF Schechter function with a Gaussian kernel, which is characterised by an r.m.s. given by the stellar mass error  (see e.g. Zucca et al.~1997; Teerikorpi~2004 and Ilbert et al.~2013 for more details). Here, we actually considered two Gaussian kernels, whose characteristic r.m.s. values correspond to the two stellar mass errors described above (one produced by the $z_{\rm phot}$ and another at fixed redshift), taking into account these errors on each galaxy on an individual basis.  We then repeated the STY maximum likelihood analysis on our real galaxy sample, adopting this convolution as the STY functional form.

\begin{figure*}
\epsscale{1.1}
\plotone{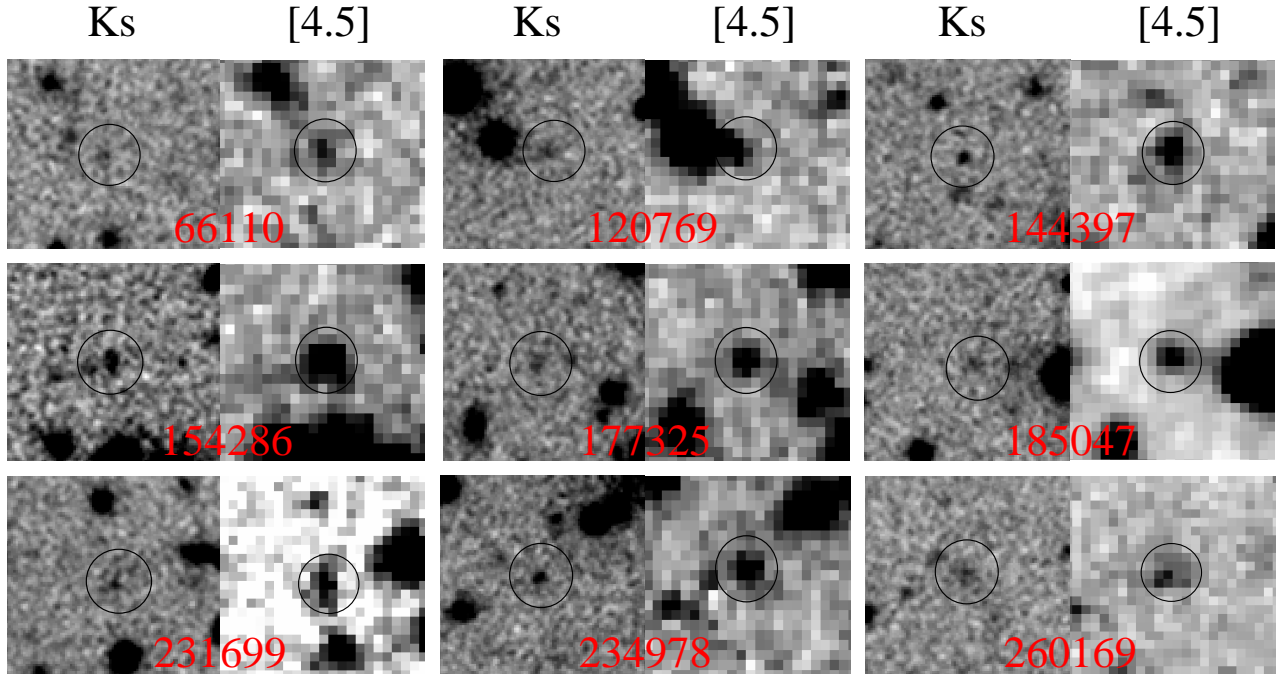}
\caption{$K_s$ and $4.5 \, \rm \mu m$ postage stamps of our $z\geq 5$  galaxy candidates. The field shown in each stamp is of $\sim 12 \times 12$~arcsec$^2$. 
\label{fig-stampszge5}}
\end{figure*}

We found that the best-fit parameters of the `real' Schechter function are somewhat different from those obtained for the  STY analysis with a plain Schechter function without error convolution (see Table \ref{tab-schech35}). This is especially the case for the resulting $\alpha$ value: for the `real' Schechter functions, we get $\alpha=1.58 \pm 0.06$ and $\alpha=1.58^{+0.12}_{-0.16}$, at $3.0 \leq z < 4.0$ and $4.0 \leq z <5.0$, respectively. These values are consistent with those derived by Grazian et al.~(2015), after a similar Eddington bias analysis. At the same time, the change we found in the derived value of $M^\ast$ is small, which implies that the resulting best Schechter curve after error deconvolution does not differ much with respect to the original Schechter curve at the highest-mass end (Figure~\ref{fig-gsmfedd45}).

This result is indicating that our originally derived GSMF is affected by the Eddington bias. After correction, it still holds that the real Schechter function $\alpha$ value is higher at $z\geq3$ than in the local Universe, but the differences are less dramatic than those resulting from the GSMF analysis taken at face value.

\section{Massive galaxy candidates at $\lowercase{z} \geq 5$}
\label{sec-massz57}

\subsection{Sample properties}
\label{sec-zgeq5}

Our $[4.5]<23$, $\ksau$ sample contains a total of 9 $z \geq 5$ galaxy candidates, out of which only one is at $z>6$ (with best $z_{\rm phot}=6.04$). As explained in Section~\ref{sec-sedz}, this is a conservative list of objects, for which both the $\chi^2_{\rm min}$ photometric redshifts and median of the $P(z)$ distribution indicate that these are very likely $z \geq 5$ galaxies. We list the coordinates and properties of these objects in Table~\ref{tab-zge5}, and show their near- and mid-IR images in Figure~\ref{fig-stampszge5}.  

\begin{figure*}
\epsscale{1.1}
\plotone{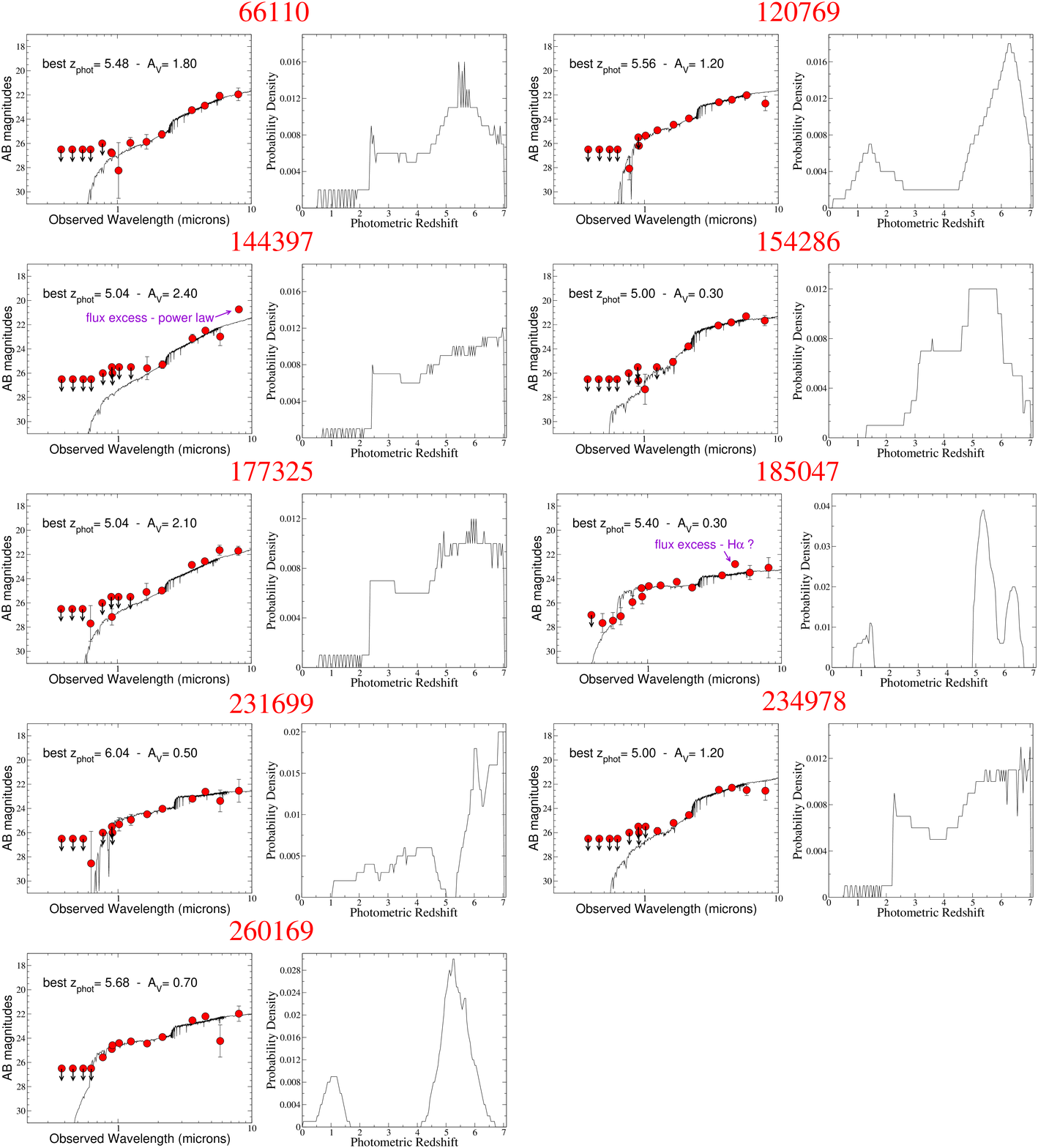}
\caption{Best SED fittings and marginalized probability density distributions $P(z)$ versus $z$  for our $z\geq 5$  galaxy candidates. For clarity, only the broad-band photometric data points are shown in the SEDs.
\label{fig-zge5exam}}
\end{figure*}

In addition, Figure~\ref{fig-zge5exam} shows the best-fitting SEDs and marginalized $P(z)$ versus $z$ distributions of these 9 galaxies. From these plots, we see that the presence of the 4000$\rm \AA$ break shifted beyond the $K_s$ band, in combination with dust extinction,  produces the near-to-mid-IR red colours of these sources. In one case, instead, the red colour is mainly produced by a $4.5 \, \rm \mu m$ flux excess, which is likely due to the presence of an emission line. For this source, our best $z_{\rm phot}$ value (obtained only with continuum SED fitting) is consistent with the H$\alpha$ emission line ($\lambda_{\rm rest}=6563 \rm \AA$) being shifted into the $4.5 \, \rm \mu m$ band. Note that, as expected, the $P(z)$ distributions have well-localized peaks when the sources are detected in many bands, but become broader when there are multiple non-detections. 

The best-fitting extinction values of our $z_{\rm phot} \geq 5$ candidates span a range between $A_V=0.30$ and $2.40$~mag. We have identified one source with an IR power-law excess (id \#144397), which, at these redshifts, indicates the presence of a very warm AGN. As before, this has been determined through the SED analysis extended to all IRAC bands, performed following the prescription of Caputi~(2013). This suggests that dust-obscured AGN have been present in the Universe since the first few billion years of cosmic time.
Indeed, another $z \geq 5$  candidate, source \#260169, is detected at $24 \, \rm \mu m$ with flux density $S_\nu(24 \, \rm \mu m)=(133 \pm 19) \, \rm \mu Jy$.   The redshift estimate of this source is one of the most secure among our $z \geq 5$ candidates -- note that it is detected in most of the near-IR and optical wavebands, except at the shortest wavelengths (Fig.~\ref{fig-zge5exam}). Hence, this $24 \, \rm \mu m$ emission is very likely due to an AGN.

On the other hand, source \#154286 may also have a (marginal) $24 \, \rm \mu m$ detection, but the exact flux density is difficult to determine precisely, as it is blended with a bright neighbour at $24 \, \rm \mu m$ (which can be seen at the bottom of the IRAC stamps shown in Fig.~\ref{fig-stampszge5}). So, for this source, there could be two possibilities: either it has an AGN mid-IR excess, as source \#260169, or the real redshift is lower than our best value quoted here ($z_{phot}=5.00$). Given the current information, we cannot decide among these two possibilities, but we decided to keep this source in our $z \geq 5$ candidate list, as it complies with all our other selection criteria. Neither this source nor any other of our $z \geq 5$ candidates are detected in the latest SCUBA2 maps of the COSMOS field (Chen et al., in preparation).

\begin{deluxetable*}{rrrccc}
\tabletypesize{\scriptsize}
\tablecaption{IRAC coordinates and properties of the $z \geq5$ galaxy candidates.  \label{tab-zge5}}
\tablehead{\colhead{ID} & \colhead{RA(J2000)} &  \colhead{DEC(J2000)} & \colhead{$z_{\rm phot}$} & \colhead{$A_V$ (mag)} & \colhead{$M_{\rm st} \, \rm (\times 10^{11} \, M_\odot)$}
}  
\startdata
66110  & 09:57:38.80 & +01:44:37.9 & $5.48^{+1.04}_{-0.92}$ & $1.80\pm0.45$ & $2.95 \pm 0.95$ \\ 
120769 & 09:57:41.32 & +01:59:41.5 & $5.56^{+1.40}_{-0.68}$ & $1.20\pm0.30$ & $1.23 \pm 0.39$ \\  
144397 & 10:00:47.88 & +02:06:09.2 & $5.04^{+1.96}_{-2.28}$ & $2.40\pm0.60$ & $1.85 \pm 0.59$ \\   
154286 & 10:00:56.69 & +02:08:49.4 & $5.00^{+1.24}_{-1.98}$ & $0.30\pm0.30$ & $3.35 \pm 1.10$ \\
177325 & 09:57:28.92 & +02:14:46.6 & $5.04^{+1.96}_{-1.80}$ & $2.10\pm0.60$ & $1.37 \pm 0.44$ \\
185047 & 10:01:56.87 & +02:16:51.2 & $5.40^{+0.80}_{-0.40}$ & $0.30\pm0.30$ & $0.63 \pm 0.20$ \\
231699 & 09:57:27.65 & +02:29:06.6 & $6.04^{+0.96}_{-0.24}$ & $0.50\pm0.30$ & $1.77 \pm 0.57$ \\
234978 & 10:02:12.64 & +02:30:01.0 & $5.00^{+1.80}_{-0.96}$ & $1.20\pm0.45$ & $2.76 \pm 0.88$ \\
260169 & 10:01:57.76 & +02:36:48.2 & $5.68^{+0.16}_{-0.72}$ & $0.70\pm0.30$ & $1.00 \pm 0.32$
\enddata
\end{deluxetable*}

\subsection{The GSMF at $z \geq 5$}
\label{sec-gsmfz5}

\subsubsection{Constraints to the GSMF high-mass end at $5 \leq z \leq 7$}

\begin{figure*}
\epsscale{1.1}
\plotone{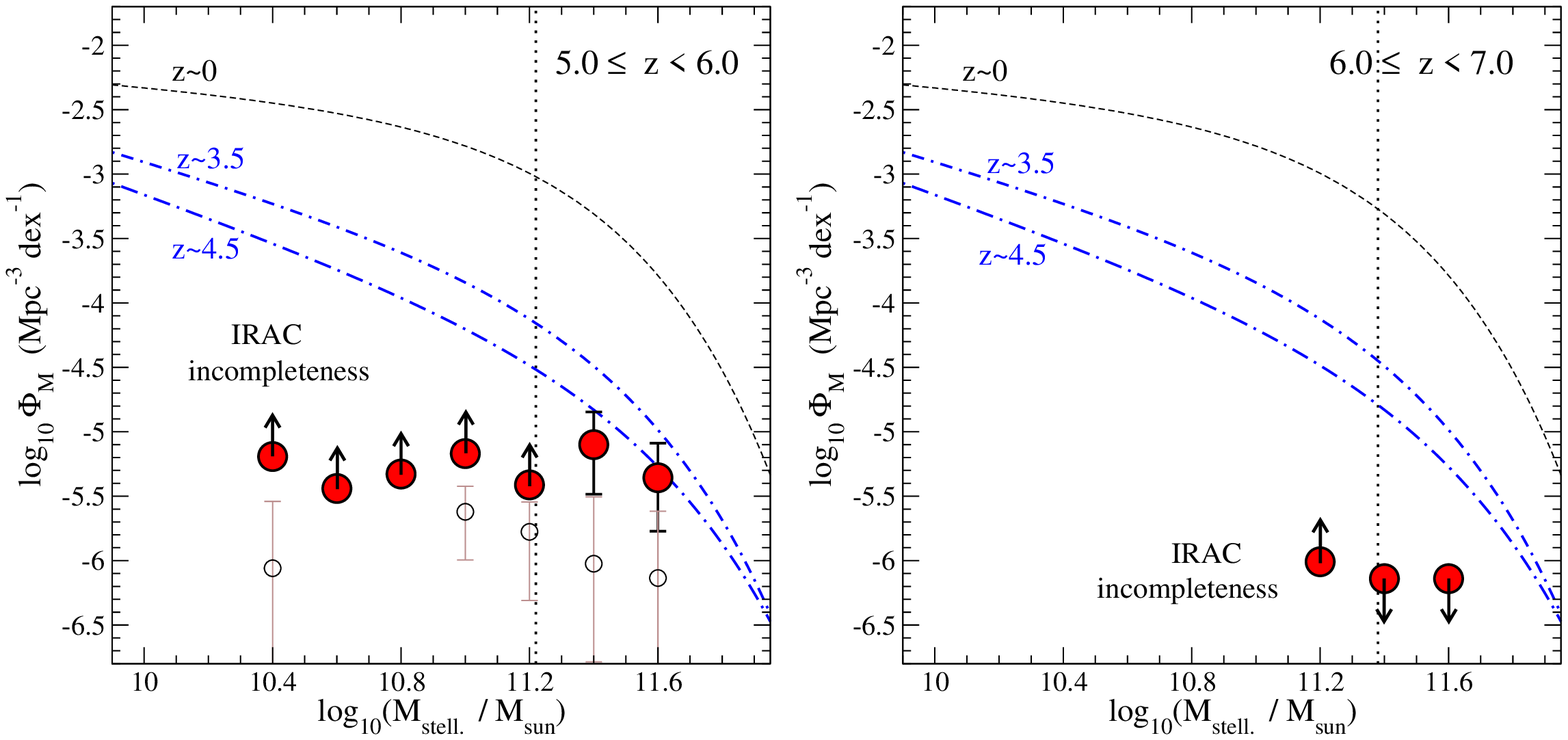}
\caption{GSMF high-mass end at $5.0 \leq z < 6.0$ ({\em left}) and $6.0 \leq z < 7.0$ (right). For the GSMF at $5.0 \leq z < 6.0$, we considered the combination of all UltraVISTA sources in the DR1 release, and our new sources with $[4.5]<23$, $\ksau$ over the UltraVISTA ultra-deep stripes. Symbols are the same as in Fig.~\ref{fig-gsmf35}. The GSMF constraints at $6.0 \leq z < 7.0$ are based on our single $z>6$ candidate (as no $z \geq 6$ galaxies were identified in the UltraVISTA DR1) and upper limits obtained from the absence of very massive galaxies in the UltraVISTA ultra-deep area. Note that with our sample selection cut at $[4.5]<23$, we only aim at constraining the highest-mass end of the GSMF at these high redshifts. 
\label{fig-gsmfz57}}
\end{figure*}

Although here we only find a small number of galaxies at $5 \leq z < 7$, it is still important to understand their contribution to the GSMF. Even the fact that we find only one galaxy at $z>6$ provides important constraints at high $z$, given our large surveyed area. Note that we only aim at constraining the highest-mass end of the GSMF at these high redshifts with our sample selection cut at $[4.5]<23$. In this sense, our study is complementary to those selecting, e.g., Lyman-break galaxies at these redshifts, which sample the intermediate-mass regime. Our possibility of investigating the GSMF highest-mass end at $z_{\rm phot} \geq 5$ is unique, thanks to the combination of the COSMOS large area and depth of the UltraVISTA survey.

\begin{deluxetable}{ccc}
\tabletypesize{\scriptsize}
\tablecaption{GSMF constraints at $5 \leq z \leq 7$.  \label{tab-vmaxz57}}
\tablehead{\colhead{$\log_{10} \rm (M/M_\odot)$} &  \multicolumn{2}{c}{$\log_{10} (\rm \Phi_M / Mpc^{-3} \, dex^{-1})$} \\
 &  $5.0 \leq z <6.0$ & $6.0 \leq z \leq 7.0$
}  
\startdata
10.40 & $> -5.19$  & $\cdots$ \\
10.60 & $> -5.44$ &  $\cdots$  \\
10.80 & $> -5.33$ &  $\cdots$  \\
11.00 & $> -5.17$ &  $\cdots$   \\  
11.20 & $> -5.41$ & $>-6.01$   \\ 
11.40 & $-5.10^{+0.25}_{-0.39}$ &  $<-6.14$   \\ 
11.60 & $-5.36^{+0.27}_{-0.42}$ &   $<-6.14$     
\enddata
\end{deluxetable}

Figure~\ref{fig-gsmfz57} shows our GSMF computed with the $V/V_{\rm max}$ method at redshifts $5 \leq z_{\rm phot} < 6$ and  $6 \leq z_{\rm phot} < 7$ (see also Table~\ref{tab-vmaxz57}). For the GSMF at $5 \leq z < 6$, we considered a combination of the UltraVISTA galaxies with $K_s^{\rm auto}< 24$ (comprising 70 galaxies from the DR1 over $\sim 1.5 \, \rm deg^2$), and the new galaxies with $[4.5]<23$, $\ksau$ studied here. At $6 \leq z_{\rm phot} < 7$, instead, we only considered our single $z>6$ galaxy in our new sample, as no robust galaxy candidate with $K_s^{\rm auto}< 24$ has been identified in the UltraVISTA DR1 at these high redshifts.   The upper limits have been computed considering the fact that we have no galaxies with $M>2 \times 10^{11} \, \rm M_\odot$ at $z \geq 6$ in the UltraVISTA ultra-deep area.

From our GSMF determination at $z_{\rm phot} \geq 5$, it may seem apparent that the highest-mass end at $M_{\rm st}>1.6-2.0 \times 10^{11} \, \rm M_\odot$ had only a modest evolution from $z\sim 5.5$ to $z\sim3.5$. However, note that the number density of such massive galaxies still rose by a factor of $\sim 4$ in a time period of only $\sim 0.8$~Gyr.

Instead, the number density of such very massive galaxies drops sharply at $z>6$. We find almost one dex difference in the number density of very massive galaxies between $z\sim6.5$ and $z\sim 5.5$. Taking into account that the cosmic time elapsed between these two redshifts is of only 0.2~Gyr, this result is striking, as we could be pinpointing quite precisely in cosmic time the moment in which the first significant population of very massive galaxies appear. We discuss this further in Section~\ref{sec-disc}.

\subsubsection{Comparison with other recent studies}

\begin{figure}
\epsscale{1.1}
\plotone{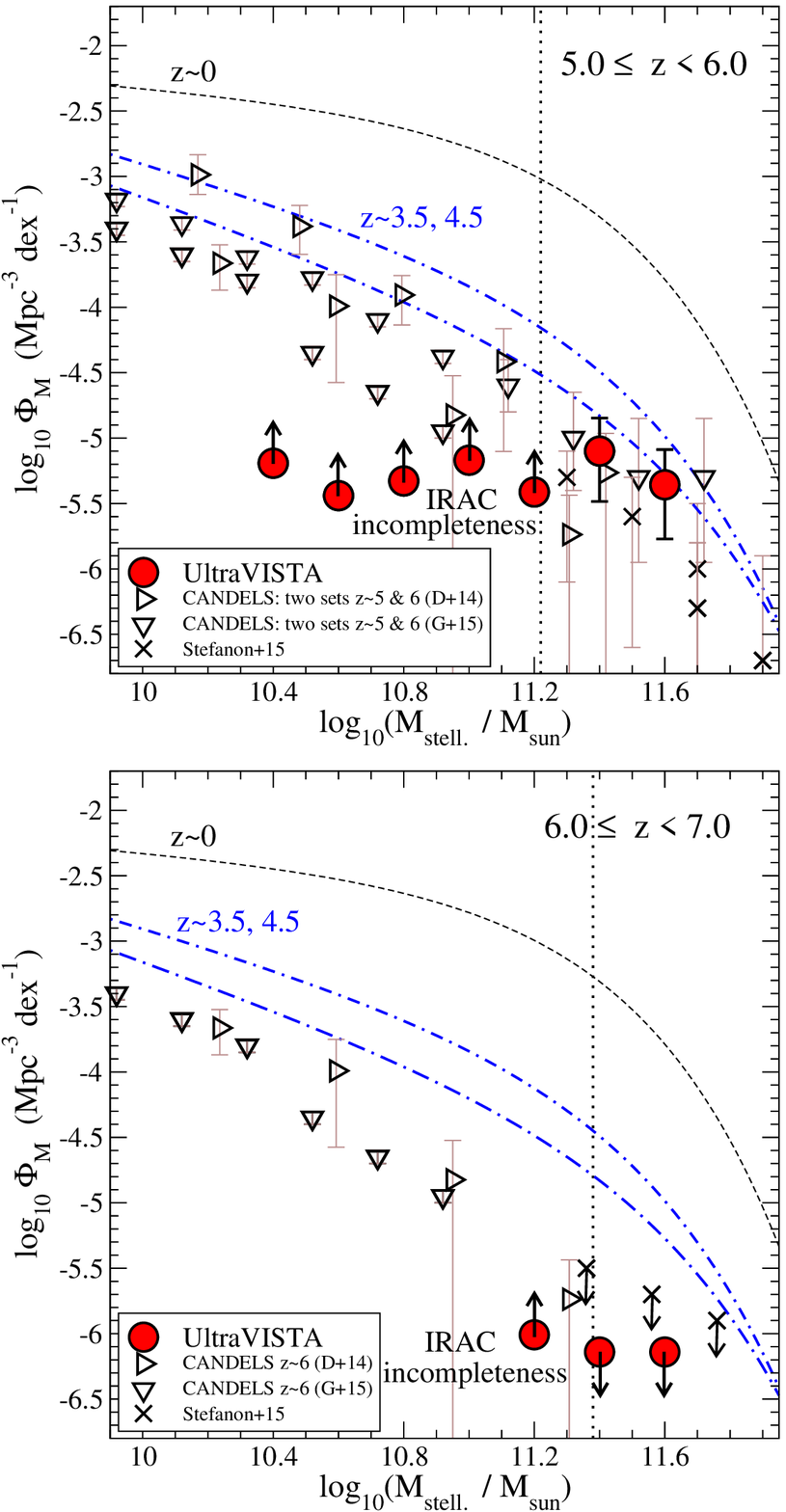}
\caption{GSMF constraints provided by our sample at $5.0 \leq z < 7.0$, in the context of the results obtained in the CANDELS survey \citep{dun14,gra15}, and the upper limits obtained by Stefanon et al.~(2015), at similar redshifts.   As in Fig.~\ref{fig-gsmf35}, the vertical dotted line indicates the stellar mass completeness limit imposed by our IRAC magnitude cut. Stellar masses correspond to a Salpeter IMF over $(0.1-100) \,\rm M_\odot$ in all cases.  \label{fig-gsmfcomp57}}
\end{figure}

In Fig.~\ref{fig-gsmfcomp57}, we show our GSMF constraints in the context of other recent results based on the study of near-/mid-IR selected galaxies. Some of these GSMF studies are based on datasets from the CANDELS survey \citep{dun14,gra15}. These authors used the ultra-deep ($H \sim 27-28$~mag) CANDELS data to select galaxies over an area that is at least nine times smaller than that covered by the UltraVISTA ultra-deep survey. The other results considered for comparison here are those obtained by Stefanon et al.~(2015), which are based on the IRAC S-COSMOS data and the UltraVISTA DR1 release.

From the results in CANDELS, we see that the GSMF covers a stellar-mass regime that is complementary to that studied here: CANDELS probes mainly the intermediate-mass regime ($M_{\rm st} \sim 10^{10}-10^{11} \, \rm M_\odot$) at $z>4$, while here we constrain the GSMF highest-mass end. Indeed, note that the CANDELS sample virtually contains no galaxy with $M_{\rm st} \gsim 10^{11} \, \rm M_\odot$ at $z>5.5$. This is very likely an effect of the small volume sampled by CANDELS. In the UltraVISTA ultra-deep stripes, we have a total of 22 galaxies with $M_{\rm st} \gsim 10^{11} \, \rm M_\odot$ at $z > 5.5$ (including all DR1 sources with $K_s^{\rm auto}< 24$ and our new $\ksau$ sources). If the distribution of these galaxies were perfectly homogeneous (i.e., if there were no cosmic variance), then at most one or two galaxies would be expected in the CANDELS field area analysed by Duncan et al.~(2014) and Grazian et al.~(2015).

In Fig.~\ref{fig-gsmfcomp57}, we also show the GSMF values and upper limits derived by Stefanon et al.~(2015). These authors have used the same S-COSMOS IRAC sample that we used here, but in combination with the UltraVISTA DR1 data release. Therefore, in this work we have additional information to that presented by Stefanon et al., which allows us to make a better estimate of the GSMF highest-mass end at $z \geq 5$. We see that our GSMF determinations are consistent with the values derived by Stefanon et al. at $5 \leq z \leq 6$ within the error bars.  At $6 \leq z \leq 7$, we place more stringent upper limits to the number density of galaxies with $M_{\rm st} \gsim 2 \times 10^{11} \, \rm M_\odot$, which are about 0.3-0.5~dex lower than their previous determination (see discussion in Section~\ref{sec-disc}).

\subsection{Discussion: galaxies with $M_{\rm st} > 10^{11} \, \rm M_\odot$ at $z \sim 6$?}
\label{sec-disc}

In this paper we find only one massive galaxy candidate at $z \geq 6$ over the UltraVISTA ultra-deep stripes area ($\sim$0.8~deg$^2$).  The best-fit SED indicates that this is a 0.1~Gyr old galaxy at  $z_{\rm phot}=6.04$ (the age of the Universe is $\sim$0.9~Gyr at $z=6$), with  extinction $A_V=0.30$. This source is not detected at $24 \, \rm \mu m$, as expected for such a distant source (unless it were an AGN). The derived stellar mass is $M_{\rm st} \approx 1.8 \times 10^{11} \, \rm M_\odot$. This single galaxy implies a minimum number density of $\sim 1.3 \times 10^{-7} \, \rm Mpc^{-3}$ for $M_{\rm st} > 10^{11} \, \rm M_\odot$ at $z\sim6$. Note that the stellar mass value is below the completeness limit imposed by our $[4.5]<23$ cut, so we should consider this number density as a lower limit.

According to the Sheth \& Tormen~(1999) formalism, the expected number density of dark matter haloes with $M>10^{12.5} \, \rm M_\odot$ is $\sim 10^{-7} \, \rm Mpc^{-3}$ at $z=6$, which is similar to the implied number density of $M_{\rm st} > 10^{11} \, \rm M_\odot$ galaxies that we obtain here. However, as our figure is a lower limit, we could plausibly expect some conflict between the observed number density of massive galaxies and the number density of dark matter haloes that can host them. Of course, identifying galaxies with haloes assumes that we know the baryon conversion fraction at high redshifts, which is in fact not known. And it is unclear whether massive haloes at those redshifts would host massive galaxies or rather multiple lower-mass galaxies (both cases are possible depending on the model parameters).

In fact, current semianalytic models basically do not predict any galaxy with $M \gsim 5 \times 10^{10} \, \rm M_\odot$ at $z=6$ (Yu Lu, private communication). However, there is no reason to conclude that the different candidates found so far in CANDELS and in this work are not real, and merely the consequence of errors in the photometric redshifts and stellar mass determinations.

Our results also set strong upper limits to the number density of galaxies with $M_{\rm st} \gsim 2 \times 10^{11} \, \rm M_\odot$ at $z>6$, by finding no robust candidate with such high stellar masses at these redshifts. Remarkably, the number density is significantly higher at $5 \leq z < 6$, which strongly suggests that the most massive galaxies in the Universe only become that massive at $z<6$. 

As an alternative, one could wonder whether there could still be very massive galaxies with $M_{\rm st} \gsim 2 \times 10^{11} \, \rm M_\odot$ at $z\sim6$, among the $\sim$1\% of $[4.5]<23$ sources that remain unidentified in the UltraVISTA DR2 survey, and/or among the sources with unknown redshifts in our current sample. A few of the latter have similar near-/mid-IR colours to our $z\sim6$ candidate (Fig.~\ref{fig-colz}), but unfortunately these colours are not conclusive for $z\sim6$ objects, as some sources at $z<3$ and a few at $3 \leq z <5$ display similar colours. Although in principle we cannot completely exclude the possibility that very massive galaxies exist at $z>6$, we note that this could only happen if these sources had significant dust extinction. With no dust, the 4000~$\rm \AA$ break of a maximal age galaxy at $z=6$ would produce $K_s-[4.5] \lsim 1$, so all the $[4.5]<23$ with these characteristics should now be identified in the UltraVISTA survey. Hence, we conclude that, unless sources with significant amounts of dust extinction exist at $z>6$, then the appearance of the most massive galaxies only happens at $5<z<6$. Future studies with ALMA and  the {\em James Webb Space Telescope} will allow us to confirm whether such very massive galaxies really exist at $z>6$, but candidates will need to be found in advance with large-area ultra-deep near-IR surveys.

\section{The evolution of the cosmic stellar mass density}
\label{sec-cstmd}

\begin{figure}
\epsscale{1.1}
\plotone{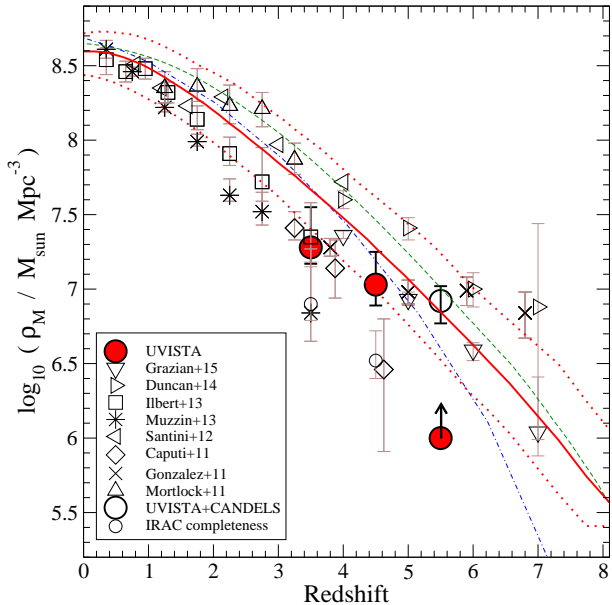}
\caption{Redshift evolution of the cosmic stellar mass density, given by our new  GSMF determinations and a compilation of recent results from the literature. At $3 \leq z <5$, our stellar mass density values have been computed by integrating the best STY  Schechter functions obtained here at stellar masses $\log_{10} (M)>10^8 \, \rm M_\odot$ (after correction for Eddington bias). At $5 \leq z < 6$, we only sum up the contributions of our galaxies in the stellar mass bins above the IRAC completeness limit (see Fig.~\ref{fig-gsmfz57}), so the stellar mass density constitutes a lower limit. The open circle indicates our best estimate of the {\em total} stellar mass density at $5 \leq z < 6$, taking into account the results of the CANDELS survey and our own, in complementary stellar-mass regimes.  We include also different model predictions:  solid and dotted lines correspond to the mean and 99\% confidence interval in Lu et al.~(2014); the dashed and dot-dashed lines correspond to mean values in Somerville et al.~(2012) and Croton et al.~(2006), respectively. All stellar mass density values in this plot refer to a Salpeter IMF over $(0.1-100) \, \rm M_\odot$.
\label{fig-stmdvsz}}
\end{figure}

We used our new GSMF determinations to set updated constraints on the cosmic stellar mass density up to $z=6$. Figure~\ref{fig-stmdvsz} shows the evolution of the cosmic stellar mass density given by our new results and a compilation of recent results from the literature. Our values at $3 \leq z <5$ have been obtained by integrating the Schechter function after correction for Eddington bias, with the parameter values given in Table~\ref{tab-schech35}, so they can be considered total values,  and we also show the values obtained by integrating the GSMF only above the stellar-mass completeness limits. At $5 \leq z \leq 6$, instead, we do not probe a wide enough stellar mass range to attempt a good Schechter function determination, so we only sum up the individual contributions of our galaxies (see Fig.~\ref{fig-gsmfz57}). Therefore, at these high redshifts, our sample provides only a lower limit of the cosmic stellar mass density. 

As the stellar mass regime that we probe at $5 \leq z \leq 6$ is complementary to that probed by CANDELS, we can can estimate the {\em total} stellar mass density estimates at these redshifts by summing up the contributions obtained here and the average of the CANDELS determinations \citep{dun14,gra15}. In this case, we only considered the galaxies in our sample above our stellar-mas completeness limit (imposed by the IRAC completeness) and the CANDELS results at lower stellar masses. The resulting total stellar mass density estimate is indicated with a large open circle in Figure~\ref{fig-stmdvsz}.

Our new stellar mass density determinations at $3 \leq z <5$ are in agreement with most recent results from the literature, within the error bars. However, at $z\sim3-4$ we cannot reproduce the high stellar mass density values found by Mortlock et al.~(2011) and Santini et al.~(2012), even within the errors. These two studies have been based on very small areas of the sky, so cosmic variance is the most likely reason for this discrepancy. On the other hand, at $z\sim4-5$, our new stellar mass density determination is significantly above the value obtained by Caputi et al.~(2011). In this case, this is mainly due to the incompleteness in the near-IR identification of high-$z$ IRAC sources (Caputi et al.~(2011) used the UDS data release 5). This effect clearly illustrates the importance of the ultra-deep near-IR data in identifying massive galaxies, and determining the total cosmic stellar mass density, particularly at $4 \lsim z \lsim 5$.

Fig.~\ref{fig-gsmfz57} also includes different model predictions of the cosmic stellar mass density evolution \citep{cro06,som12,lu14}. These models, and particularly the latest ones, reproduce reasonably well the stellar mass density values up to $z\sim7$, as they are dominated by the contribution of intermediate and low stellar mass galaxies. However, they cannot reproduce the presence of massive galaxies with $M_{\rm st} \gsim 5 \times 10^{10} \, \rm M_\odot$ up to $z\sim6$ (Yu Lu, private communication).

\section{Summary and Conclusions}
\label{sec-conc}

In this paper we have studied a sample of 574 {\em Spitzer} IRAC bright ($[4.5]<23$), near-IR faint ($\ksau$) galaxies over $\sim$0.8~deg$^2$ of the UltraVISTA  ultra-deep COSMOS field. This is the first time that the study of such galaxies can be conducted over a large area of the sky, and it is becoming possible thanks to the unique combination of area and depth that is being achieved by the UltraVISTA survey. Our galaxy sample constitutes a small ($<1\%$) fraction of the overall $[4.5]<23$ galaxy population. However,  in this paper we have shown  that these  galaxies, previously unidentified in statistically large samples, provide a significant contribution to the total population of massive galaxies at high redshifts.

Indeed, from the SED analysis of our  $[4.5]<23$,  $\ksau$ galaxies, we have determined that their redshift distribution peaks at redshift $z\sim2.5-3.0$, and $\sim32\%$ of them lie at $3 \leq z \leq 6.04$. We found that colours $H-[4.5]>4$ almost exclusively select galaxies at $3 \leq z <5$, consistently with the results of Caputi et al.~(2012).

We analysed the contribution of our $z \geq 3$ galaxies to the  GSMF high-mass end at high redshifts. We found that our galaxies make a very minor contribution to the GSMF at $3 \leq z <4$, previously determined with $K_s^{\rm auto}< 24$ galaxies. Instead, they have a more significant role within the $4 \leq z <5$ GSMF, accounting for $\gsim 50\%$ of the galaxies with stellar masses $M_{\rm st} \gsim 6 \times 10^{10} \, \rm M_\odot$. {\em We conclude that considering these IRAC bright, near-IR faint galaxies at  $4 \leq z <5$ is extremely important to properly sample the high-mass end of the GSMF at these redshifts}.  In agreement with previous works, we confirm that the GSMF Schechter function parameter $\alpha$ is significantly higher at $z \gsim 3$ than in the local Universe. However, the difference appears to be less dramatic than previously found, after correcting for the effect of Eddington bias.

Our results indicate that some very massive galaxies are present since the Universe was only a billion years old. In the $\sim 0.8 \, \rm Gyr$ of elapsed time between redshifts $z\sim 5.5$ and 3.5, the GSMF highest-mass end had a non-negligible evolution. Quantitatively, the number density of $M_{\rm st} \sim 2 \times 10^{11} \, \rm M_\odot$ galaxies rose by a factor of about four between these redshifts. The number density at $z\sim3.5$ increased by another factor of ten later by $z\sim1.5 - 2.0$, in the two billion years encompassing the star formation activity peak of the Universe \citep{hop06,beh13}. Finally, that number density increased by another factor of about four until reaching the present value.  So, in conclusion, the most massive galaxies were formed quite effectively after the first billion years: almost as effectively as during the peak activity epoch, and much more effectively than over the past ten billion years of cosmic time.

The presence of very massive ($M_{\rm st} \gsim 2 \times 10^{11} \, \rm M_\odot$) galaxies in our sample at $5 \leq z < 6$, and virtual absence at $6 \leq z < 7$,  provide a strong constraint on the evolution of the GSMF highest-mass end, which suggests that the appearance of such massive galaxies took place in the few hundred million years of elapsed time between $z\sim6$ and $z\sim5$. This kind of constraint cannot be obtained from small area surveys like CANDELS \citep{gro11,koe11}, even when the datasets are deeper. Hence, we conclude that {\em wide-area galaxy surveys are necessary to sample this very massive galaxy population at high $z$}. The combination of area and depth of UltraVISTA is currently unique for this purpose.

The only alternative to this conclusion is that, among the $[4.5]<23$ galaxies that remain unidentified and/or those which have no redshift determination in our current sample, there is a population of very massive galaxies at $z>6$ that are significantly dust obscured.  One possible candidate for such galaxies was discussed by Caputi et al.~(2012)  in one of the CANDELS fields, but the level of dust obscuration ($A_V=0.90$~mag) is atypical,  given our current knowledge of galaxies in the early Universe (but see Oesch et al.~2015). Further studies in other fields, as well as future follow up with {\em JWST} and ALMA, are necessary to confirm whether such sources exist at $z>6$. 

In any case, a substantial fraction of the still unidentified IRAC $[4.5]<23$ sources are more likely dust-obscured massive galaxies ($M_{\rm st} \gsim  5 \times 10^{10} \, \rm M_\odot$) at $4<z<6$. This is suggested by the colour-colour diagram shown in Fig.~\ref{fig-colz}  and our GSMF determinations. Further analysis of this problem after the UltraVISTA completion, which will achieve near-IR photometry $\sim 0.5$ mag deeper than the current DR2 release, will help us elucidate whether we are still missing a significant amount of massive galaxies at high $z$.

\acknowledgments

Based on data products from observations made with ESO Telescopes at the La Silla Paranal Observatory under ESO program ID 179.A-2005 and on data products produced by TERAPIX and the Cambridge Astronomy Survey Unit on behalf of the UltraVISTA consortium. Based on observations carried out with the {\em Spitzer Space Telescope}, which is operated by the Jet Propulsion Laboratory, California Institute of Technology under a contract with NASA;  the NASA/ESA {\em Hubble Space Telescope}, obtained and archived at the Space Telescope Science Institute; and the Subaru Telescope, which is operated by the National Astronomical Observatory of Japan.  This research has made use of the NASA/IPAC Infrared Science Archive, which is operated by the Jet Propulsion Laboratory, California Institute of Technology, under contract with NASA. 

OI acknowledges funding from the French Agence Nationale de la Recherche (ANR) for the project SAGACE. BMJ and JPUF acknowledge
support from the ERC-StG grant EGGS-278202. The Dark Cosmology Centre is funded by the DNRF. We thank Kenneth Duncan, Yu Lu and Paola Santini for providing us data in electronic format;  Rebecca Bowler for performing photometric tests; and Yu Lu for useful discussions. We also thank an anonymous referee for a constructive report.

\end{document}